\documentclass[onecolumn]{IEEEtran}
\usepackage{amssymb,amsmath,amsfonts,amsthm}
\usepackage{color}
\usepackage{cases}
\usepackage{standalone}
\usepackage{mathtools}
\usepackage{float}
\usepackage{cite}
\usepackage{suffix}
\usepackage{dblfloatfix}
\restylefloat{table}
\usepackage{amsfonts}
\usepackage{booktabs}
\usepackage{siunitx}
\usepackage{mathrsfs}
\usepackage{relsize}
\usepackage{graphicx}		
\usepackage{epstopdf}	
\usepackage{pgfplots}
\usepackage{pgfplotstable}
\usepackage{tikzscale} 
\usepackage{tikz}
\usetikzlibrary{shapes}
\usepackage{algorithm}
\pgfplotsset{compat=newest}	
\pgfplotsset{plot coordinates/math parser=false}
\newlength\figureheight 
\newlength\figurewidth
\usepackage{tabularx}
\usepackage{enumerate}
\usepackage{mathrsfs}
\usepackage{amsbsy}
\usepackage{units}
\usepackage{bm}

\newcommand*\diff{\mathop{}\mathrm{d}}
\usepackage{textcomp}
\usepackage{placeins}
\usepackage{bbm}
\usepackage{textgreek}
\newcommand{\ROME}[1]{%
\textup{\uppercase\expandafter{\romannumeral#1}}%
}
\usepackage{setspace}
\DeclarePairedDelimiterX\MeijerM[3]{\lparen}{\rparen}%
{\begin{smallmatrix}#1 \\ #2\end{smallmatrix}\delimsize\vert\,#3}
\newcommand\MeijerG[8][]{%
\mathcal{G}^{\,#2,#3}_{#4,#5}\MeijerM[#1]{#6}{#7}{#8}}
\WithSuffix\newcommand\MeijerG*[7]{%
\mathcal{G}^{\,#1,#2}_{#3,#4}\MeijerM*{#5}{#6}{#7}}
\newtheorem{thm}{Theorem}
\newtheorem{lem}{Lemma}
\newtheorem{rem}{Remark}
\newtheorem{cor}{Corollary} 
\newtheorem{asm}{Assumption} 
\newtheorem{prp}{Proposition} 
\DeclareMathOperator{\arccot}{arccot}
\DeclareMathSizes{12}{10}{8}{8}

\DeclareMathOperator*{\argmax}{arg\,max}

\makeatletter
\let\MYcaption\@makecaption
\makeatother
\usepackage[font=footnotesize]{subcaption}
\makeatletter
\let\@makecaption\MYcaption
\makeatother
\allowdisplaybreaks[4]
  
\begin{document}

\title{Massive MIMO-Enabled\\ Full-Duplex Cellular Networks} 
\author{
{\large Arman Shojaeifard, \textit{Member,~IEEE}, Kai-Kit Wong, \textit{Fellow,~IEEE},\\ Marco Di Renzo, \textit{Senior Member,~IEEE}, Gan Zheng, \textit{Senior Member,~IEEE},\\ Khairi Ashour Hamdi, \textit{Senior Member,~IEEE}, Jie Tang, \textit{Member,~IEEE}}
\thanks{A. Shojaeifard and K.-K Wong are with the Communications and Information Systems Research Group, Department of Electronic and Electrical Engineering, University College London, London, United Kingdom (e-mail: a.shojaeifard@ucl.ac.uk; kai-kit.wong@ucl.ac.uk). \par M. Di Renzo is with the Laboratoire des Signaux et Syst\`emes, CNRS, CentraleSup\'elec, Univ Paris Sud, Universit\'e Paris-Saclay, Gif-sur-Yvette, France (e-mail: marco.direnzo@l2s.centralesupelec.fr). \par G. Zheng is with the Signal Processing and Networks Research Group, Wolfson School of Mechanical, Electrical and Manufacturing Engineering, Loughborough University, Loughborough, United Kingdom (e-mail: g.zheng@lboro.ac.uk). \par K. A. Hamdi is with the Microwave and Communication Systems Research Group, School of Electrical and Electronic Engineering, University of Manchester, Manchester, United Kingdom (e-mail: k.hamdi@manchester.ac.uk). \par J. Tang is with the Smart Information Processing Centre, School of Electronic and Information Engineering, South China University of Technology, Guangzhou, China (e-mail: eejtang@scut.edu.cn). \par This work was supported by the Engineering and Physical Sciences Research Council (EPSRC) under grant EP/N008219/1.}
}
\pagestyle{empty}

\maketitle

\begin{abstract}

We provide a theoretical framework for the study of massive multiple-input multiple-output (MIMO)-enabled full-duplex (FD) cellular networks in which the residual self-interference (SI) channels follow the Rician distribution and other channels are Rayleigh distributed. To facilitate bi-directional wireless functionality, we adopt (i) a downlink (DL) linear zero-forcing with self-interference-nulling (ZF-SIN) precoding scheme at the FD base stations (BSs), and (ii) an uplink (UL) self-interference-aware (SIA) fractional power control mechanism at the FD mobile terminals (MTs). Linear ZF receivers are further utilized for signal detection in the UL. The results indicate that the UL rate bottleneck in the FD baseline single-antenna system can be overcome via exploiting massive MIMO. On the other hand, the findings may be viewed as a reality-check, since we show that, under state-of-the-art system parameters, the spectral efficiency (SE) gain of FD massive MIMO over its half-duplex (HD) counterpart largely depends on the SI cancellation capability of the MTs. In addition, the anticipated two-fold increase in SE is shown to be only achievable with an infinitely large number of antennas.

\end{abstract}


\section{Introduction}

The fifth-generation mobile network (5G) is expected to roll out from 2018 onwards as a remedy for tackling the existing capacity crunch \cite{6736746}. A key 5G technology is massive multiple-input multiple-output (MIMO), or large scale antenna system (LSAS), where the base stations (BSs) equipped with hundreds of antennas simultaneously communicate with multiple mobile terminals (MTs) \cite{6375940}. Massive MIMO, via spatial-multiplexing and directing power intently, can greatly outperform the state-of-the-art cellular standards jointly in terms of spectral efficiency (SE) and energy efficiency (EE) \cite{7401108}, \cite{DBLP:journals/corr/GoyalGMP16}. Moreover, under increasingly scarce spectrum, the transcieving of information over the same radio-frequency (RF) resources, i.e. full-duplex (FD) mode \cite{6353396}, has become a topic of interest for 5G and beyond \cite{6702851}, \cite{7263346}. In theory, FD can double the sum-rate of half-duplex (HD) systems, where orthogonal RF partitioning is typically employed to avoid the over-powering self-interference (SI). In practice, however, the FD over HD SE gain predominantly depends on the SI cancellation capability. 

There has been major breakthroughs in SI cancellation using any combination of (i) spatial/angular isolation and (ii) subtraction in digital/analog domains \cite{7041186}. FD, beyond point-to-point, remains in its infancy. In particular, the introduction of cross-mode interference (CI) between the downlink (DL) and the uplink (UL), as well as SI, significantly increases the complexity for FD cellular setups. Many relevant works have emerged recently, including FD for small-cell (SC) \cite{6847175,DBLP:journals/corr/AtzeniK15}, relay (RL) \cite{6963403,7053957,7403938}, cloud radio access network (CRAN) \cite{DBLP:journals/corr/MohammadiST16}, and heterogeneous cellular network (HCN) \cite{DBLP:journals/corr/AlAmmouriEAA15,2016arXiv160400588S,7105653}. A general consensus from early results is that the FD over HD SE gain mostly arises in the DL and that the UL is the main performance bottleneck. The authors in \cite{2016arXiv160402602E} have shown that bi-directional cellular systems with baseline single-input single-output (SISO), achieve double the DL rate at the cost of more than a thousand-fold reduction in the UL rate. A potential strategy for tackling this limitation is to exploit the large degrees of freedom (DoF) in massive MIMO for better resilience against SI and CI \cite{6832464}, \cite{7105646}.

In \cite{6810573}, the authors considered a FD BS with large scale antenna array serving multiple HD single-antenna MTs, and proposed a linear extended zero-forcing (ZF) precoder to suppress SI at the receiving antennas, subject to perfect channel state information (CSI). The sum-rate with FD mode was almost doubled versus that in the HD case, and the optimal ratio of transmit over receive antennas was found to be approximately $3$ \cite{6810573}. The extension to a FD BS with large scale antenna array serving HD multi-antenna MTs using linear block diagonalization (BD) beamforming was provided in \cite{7428973}. With asymptotically large number of antennas, the optimal ratio of transmit over receive antennas was shown to converge to the ratio of DL over UL streams. In \cite{DBLP:journals/corr/LimHC15}, under a setup similar to that in \cite{6810573}, the FD sum-rate with different spatial isolation and SI subtraction techniques was compared considering both perfect and imperfect CSI. On the other hand, non-linear transceiver design for maximizing the sum-rate in FD multi-user MIMO has been studied in \cite{6983622} and \cite{7492210}. The works highlighted considered a single-cell network. Intuitively, the introduction of SI and CI has huge implications in a FD multi-cell multi-user MIMO paradigm necessitating rigorous investigation.    
 
Multi-cell studies of FD MIMO cellular networks have also been recently reported. In \cite{BaiS16}, the DL and UL ergodic sum-rates in a deterministic FD multi-cell multi-user MIMO setup were characterized using conjugate-beamforming (CB). In addition, the throughput performance in large-scale distributed FD MIMO systems with BS cooperation (i.e., network MIMO) was investigated in \cite{7218556}. The authors utilized interference-alignment (IA) for tackling CI and characterized the FD multiplexing gain in closed-form. Moreover, in \cite{AtzeniK16a}, success probability expressions in cellular systems with FD relaying BSs under different beamforming and interference-cancellation strategies were characterized using stochastic geometry theory. On the other hand, the design and analysis of randomly-deployed FD cellular networks with directional antennas was provided in \cite{psomas16}. In particular, the authors derived analytical expressions for the DL and UL coverage probabilities under passive SI suppression and fixed power control. 
 
Motivated by the above, a theoretical stochastic geometry-based framework for the study of FD massive MIMO cellular networks using the Poisson point process (PPP)-based abstraction model is presented. We jointly consider the DL and UL of a multi-cell FD system where multi-antenna BSs and MTs are transceiving over the same RF resources. We adopt the Rician distribution for the residual SI fading channels, thus capturing performance under generalized SI cancellation capability. All other fading channels are modeled using the Rayleigh distribution. In the DL, we devise a linear zero-forcing with self-interference-nulling (ZF-SIN) precoder to jointly suppress residual SI and multi-user interference under perfect CSI. On the other hand, we propose a self-interference-aware (SIA) fractional power control mechanism at the FD MTs to keep SI below a certain level. We derive the distributions, and in certain cases higher-order statistics of the transmit power. In the UL, for signal detection at the FD massive MIMO BSs, linear ZF receivers are utilized for suppressing multi-user interference. 

The signals distributions are derived under the linear processes described above. We characterize the DL and UL SEs using a moment-generating-function (m.g.f.)-based approach and derive the signals conditional statistics in closed-form. The proposed framework can be readily used to obtain the exact SE via three-fold integrals, versus (i) the manifold integrals involved in the direct capacity evaluation approach, and (ii) the highly resource-intensive Monte-Carlo (MC) simulations \cite{GeoffreyFD}. We further reduce the computational complexity in certain special cases. In particular, we adopt the proposed framework to study the FD versus HD cellular network performance. With baseline SISO, we derive a tight bounded closed-form function of the FD over HD SE gain and show its optimal point occurs in the ratio of the BS and MT transmit powers being equal to one. With massive MIMO, we derive single-integral expressions for the FD and HD SEs and hence utilize non-linear curve fitting to develop a closed-form approximation of the corresponding FD over HD SE gain as a function of the number of antennas and users.       

The validity of the theoretical findings is confirmed using MC simulations. The results highlight that the key features of massive MIMO, in terms of high transmit/receive array gain and lower BS/MT transmit power, allow for achieving significant performance gains over other FD multi-cell setups. On the other hand, the SE gain of FD over HD massive MIMO cellular network with finite antenna arrays was shown to be predominantly dependent on the SI cancellation capability of the MTs. In addition, the corresponding sum-rate gain was shown to increase logarithmically in the antenna array size, with the anticipated two-fold increase in SE only achieved as the number of antennas tends to be infinitely large.

\textit{Organization:} The FD massive MIMO cellular network is described in Section \ROME{2}. The different signals are modeled in Section \ROME{3}. The power control mechanism is introduced in Section \ROME{4}. The SE analysis is provided in Section \ROME{5}, followed by a comparison of FD over HD in Section \ROME{6}. Numerical results are provided in Section \ROME{7}, and finally, conclusions are drawn in Section \nolinebreak[4] \ROME{8}. 

\textit{Notation:} $\boldsymbol{X}$ is a matrix with ($i,j$)-th entry $\{ \boldsymbol{X} \}_{i,j}$; $\boldsymbol{x}$ is a vector with $k$-th element $\{ \boldsymbol{x} \}_{k}$; $T$, $\dag$, and $+$ are the transpose, Hermitian, and pseudo-inverse operations; $\mathbb{E}_{x}\{.\}$ is the expectation; $\mathcal{F}_{x}(.)$ is the cumulative density function (c.d.f.); $\mathcal{P}_{x}(.)$ is the probability density function (p.d.f.); $\mathscr{P}(x)$ is the probability; $\mathcal{M}_{x}(.)$ is the m.g.f.; $| x |$ is the modulus; $\| \boldsymbol{x} \|$ and $\| \boldsymbol{X} \|$ are the Euclidean and Frobenius norms; $\mathbf{I}_{(.)}$ is the identity matrix; $\text{Null}(.)$ is a nullspace; $\mathcal{H}(.)$ is the Heaviside step function; $\delta(.)$ is the Delta function;
 $\Gamma(.)$ and $\Gamma(.,.)$ are the Gamma and incomplete (upper) Gamma functions; $\mathcal{L}_{n}(.)$ is the Laguerre polynomial; $\text{Ei}(.)$, $\mathscr{S}(.)$, and $\mathscr{C}(.)$ are the exponential, Sine, and Cosine integral functions; $\text{erfi}(.)$ is the imaginary error function; $Q_{m} (.,.)$ is the Marcum $Q$-function; $_2F_1(.,.;.;.)$, $_2\tilde{F}_1(.,.;.;.)$, $_0F_1\left(;.;.\right)$, $_0\tilde{F}_1\left(;.;.\right)$, $_pF_q(.;.;.)$, $_p\tilde{F}_q(.;.;.)$ are the Gauss, Regularized Gauss, confluent, Regularized confluent, generalized, and Regularized generalized hypergeometric functions; and $\MeijerG*{m}{n}{p}{q}{.}{.}{(.)}$ is the Meijer-G function, respectively.

\section{Modeling Preliminaries}

Consider a FD cellular setup with BSs and MTs respectively deployed on the Euclidean grid according to independent stationary PPPs $\Phi^{(d)}$ and $\Phi^{(u)}$ with spatial densities $\lambda^{(d)}$ and $\lambda^{(u)}$. Let $l$ and $k$ denote the locations of the $l$-th BS and the $k$-th MT, respectively. Their respective Euclidean distance is therefore $d_{l,k} = \| l - k \|$. The BSs are assumed to be equipped with $N_{t}$ transmit and $N_{r}$ receive antennas ($N_{t} + N_{r}$ RF chains), respectively. The MTs are in turn assumed to be equipped with single transmit/receive antennas (two RF chains). In the DL, the BSs simultaneously serve $\mathcal{U}$ MTs per resource block using linear transmit precoding. In the UL, the scheduled MTs simultaneously transmit to their serving massive MIMO BS per resource block \cite{7544500}. Linear receive filters are then used for UL signal detection. We assume the condition $\mathcal{U} \leq \min ( N_{t} , N_{r} )$ holds, thus scheduling is not necessary here. 

Let $\boldsymbol{g}_{l,k} \in \mathcal{C}^{1 \times N_{t}}$, $\boldsymbol{G}_{l,j} \in \mathcal{C}^{N_{r} \times N_{t}}$, and $\boldsymbol{G}_{l,l} \in \mathcal{C}^{N_{r} \times N_{t}}$ denote the channel from the $l$-BS to the $k$-th MT, the channel from the $l$-th BS to the $j$-th BS, and the residual SI channel at the $l$-th BS, respectively. Moreover, $\boldsymbol{h}_{k,l} \in \mathcal{C}^{N_{r} \times 1}$, $h_{k,i}$, and $h_{k,k}$ are respectively the channel from the $k$-th MT to the $l$-th BS, the channel from the $k$-th MT to the $i$-th MT, and the residual SI channel at the $k$-th MT. The residual SI (hereafter, refered to as SI) channels are subject to Rician fading with independent and identically-distributed (i.i.d.) elements drawn from $\mathcal{C} \mathcal{N} ( \mu , \upsilon^2 )$. All other channels are modeled using Rayleigh fading with i.i.d. elements drawn from $\mathcal{C} \mathcal{N} \left( 0 , 1 \right)$. We use the unbounded distance-dependent path-loss model with exponent $\beta$ ($> 2$). CSI in time-division duplex (TDD)-based massive MIMO systems can be acquired based on channel reciprocity through UL training. In this work, we assume sophisticated channel estimation algorithms with sufficient training information are used to obtain perfect CSI \cite{6891254}. 
 
By invoking the Campbell-Mecke theorem \cite{CMTheorem}, the DL analysis is carried out for an arbitrary MT $o$ assumed to be located at the center. We consider a cellular association strategy where the reference MT is exclusively served by a BS $b$ which provides the greatest received signal power. For homogeneous deployments, this is equivalent to the cell selection approach based on the closest distance
$b = \argmax \left( d^{- \beta}_{l,o} \right), \; \forall l \in \Phi^{(d)}$ \cite{6918448}. The UL analysis, on the other hand, is carried out for the reference MT at its serving BS. The reference transceiver distance p.d.f. is given by $\mathcal{P}_{d_{b,o}} \left( r \right) = 2 \pi \lambda^{(d)} r \exp \left( - \pi \lambda^{(d)} r^{2} \right)$. It should be noted that the alternative decoupling approach for cellular association \cite{7432156} results in the loss of channel reciprocity in massive MIMO systems.   

\section{Signals Definitions}

Let $\boldsymbol{G}_{l} = [\boldsymbol{g}^{T}_{l,k}]^{T}_{1 \leq k \leq \mathcal{U}} \in \mathcal{C}^{\mathcal{U} \times N_{t}}$ denote the combined DL channels from the $l$-th BS to its $\mathcal{U}$ MTs. We use $\boldsymbol{s}_{l} = [s_{l,k}]^{T}_{1 \leq k \leq \mathcal{U}} \in \mathcal{C}^{\mathcal{U} \times 1}$, $\mathbb{E} \left\{ | s_{l,k} |^2 \right\} = 1$, to denote the DL complex symbol vector from the $l$-th BS to its $\mathcal{U}$ MTs. Here, we consider the case where each BS equally allocates its total transmit power $p^{(d)}$ among its $\mathcal{U}$ MTs. The normalized precoding matrix at the $l$-th BS is $\boldsymbol{V}_{l} = [\boldsymbol{v}_{l,k}]_{1 \leq k \leq \mathcal{U}} \in \mathcal{C}^{N_{t} \times \mathcal{U}}$, $\mathbb{E} \left\{ \| \boldsymbol{v}_{l.k} \|^2 \right\} = 1$. The DL received signal can be represented as
\begin{align}
y_{d} & = \underbrace{\sqrt{\tfrac{p^{(d)}}{\mathcal{U}}} d_{b,o}^{- \frac{\beta}{2}} \boldsymbol{g}_{b,o} \boldsymbol{v}_{b,o} s_{b,o}}_{d \, (\text{intended signal})} + \underbrace{\sqrt{\tfrac{p^{(d)}}{\mathcal{U}}} d_{b,o}^{- \frac{\beta}{2}} \boldsymbol{g}_{b,o} \sum_{k \in \Phi^{(u)}_{b} \setminus \{ o \} } \boldsymbol{v}_{b,k} s_{b,k}}_{dd \, (\text{multi-user interference})} + \underbrace{ \sqrt{\tfrac{p^{(d)}}{\mathcal{U}}} \sum_{l \in \Phi^{(d)} \setminus \{ b \} } d_{l,o}^{- \frac{\beta}{2}} \boldsymbol{g}_{l,o} \boldsymbol{V}_{l} \boldsymbol{s}_{l} }_{d,d \, (\text{inter-cell interference})} \nonumber \\ & + \underbrace{\sum_{k \in \Phi^{(u)}_{l},l \in \Phi^{(d)} \setminus \{ o,b \} } \sqrt{p^{(u)}_{k,l}} d_{k,o}^{- \frac{\beta}{2}} h_{k,o} s_{k,l} }_{u,d \, (\text{cross-mode interference})} + \underbrace{ \sqrt{p^{(u)}_{o,b}} h_{o,o} s_{o,b}}_{si,d \, (\text{self-interference})} + \underbrace{\eta_{o}}_{\text{noise}}
\end{align}
where $\Phi^{(u)}_{l}$ ($\subset \Phi^{(u)}$) is the set of scheduled MTs in the cell of BS $l$, $p^{(u)}_{k,l}$ is the $k$-th scheduled MT transmit power for sending $s_{k,l}$ to BS $l$, and $\eta_{o}$ is the complex additive white Gaussian noise (AWGN) with mean zero and variance $\sigma^{2}_{d}$, respectively. 

Next, let $\boldsymbol{H}_{l} = [\boldsymbol{h}_{k,l}]_{1 \leq k \leq \mathcal{U}} \in \mathcal{C}^{N_{r} \times \mathcal{U}}$ represent the compound UL channel matrix at the $l$-BS from its $\mathcal{U}$ scheduled MTs. The linear receiver filter at the $l$-th BS is denoted using $\boldsymbol{W}_{l} = [ \boldsymbol{w}^{T}_{k,l} ]^{T}_{1 \leq k \leq \mathcal{U}} \in \mathcal{C}^{\mathcal{U} \times N_{r}}$, $\mathbb{E} \left\{ \| \boldsymbol{w}_{k,l} \|^{2} \right\} = 1$. The corresponding post-processing UL signal can be written as
\begin{align}
y_{u} & = \underbrace{\sqrt{p^{(u)}_{o,b}} d_{o,b}^{- \frac{\beta}{2}} \boldsymbol{w}^{T}_{o,b} \boldsymbol{h}_{o,b} s_{o,b}}_{u \, (\text{intended signal})} + \underbrace{\sum_{k \in \Phi^{(u)}_{b} \setminus \{ o \} } \sqrt{p^{(u)}_{k,b}} d_{k,b}^{- \frac{\beta}{2}} \boldsymbol{w}^{T}_{o,b} \boldsymbol{h}_{k,b} s_{k,b}}_{uu \, (\text{multi-user interference})} + \underbrace{\sum_{k \in \Phi^{(u)}_{l},l \in \Phi^{(d)} \setminus \{ b \}} \sqrt{p^{(u)}_{k,l}} d_{k,b}^{- \frac{\beta}{2}} \boldsymbol{w}^{T}_{o,b} \boldsymbol{h}_{k,b} s_{k,l} }_{u,u \, (\text{inter-cell interference})} \nonumber \\ & + \underbrace{\sqrt{\tfrac{p^{(d)}}{\mathcal{U}}} \sum_{l \in \Phi^{(d)} \setminus \{ b \} } d_{l,b}^{- \frac{\beta}{2}} \boldsymbol{w}^{T}_{o,b} \boldsymbol{G}_{l,b} \boldsymbol{V}_{l} \boldsymbol{s}_{l} }_{d,u \, (\text{cross-mode interference})} + \underbrace{\sqrt{\tfrac{p^{(d)}}{\mathcal{U}}} \boldsymbol{w}^{T}_{o,b} \boldsymbol{G}_{b,b} \boldsymbol{V}_{b} \boldsymbol{s}_{b} }_{si,u \, (\text{self-interference})} + \underbrace{\boldsymbol{w}^{T}_{o,b} \boldsymbol{\eta}_{b}}_{\text{noise}}
\end{align}
where $\boldsymbol{\eta}_{b} \in \mathcal{C}^{N_{r} \times 1}$ is the AWGN vector with mean zero and covariance matrix $\sigma^{2}_{u} \mathbf{I}_{N_{r}}$. 

It is important to note that the set of scheduled MTs is \textit{strictly not} an indpendent process due to the spatial dependencies arising from (i) the cellular association strategy, and (ii) the constraint of each BS serving multiple MTs per resource block \cite{6516885,6919997}. For the sake of mathematical tractability, in the same spirit as in \cite{7448861}, we invoke the following assumption.

\begin{asm}
\label{asm:UEsProcess}
The set of scheduled MTs, conditioned on the spatial constraints imposed by the cellular association strategy and the number of MTs being served by each BS per resource block, is modeled as an independent stationary PPP with density $\lambda^{(u)}$.
\end{asm}
 
\begin{prp} In the DL, we adopt a linear ZF-SIN precoder where the transmit antenna array (conditioned on $N_{t} \geq N_{r} + \mathcal{U}$) is utilized to jointly suppress SI and multi-user interference at the receiving antennas. This is achieved at the BS $l$ by setting the columns of $\boldsymbol{V}_{l}$ equal to the normalized columns of $\boldsymbol{\hat{G}}^{+}_{l} = \boldsymbol{\hat{G}}^{\dag}_{l} ( \boldsymbol{\hat{G}}_{l} \boldsymbol{\hat{G}}^{\dag}_{l} )^{-1} = [\boldsymbol{\hat{g}}_{l,k}]_{1 \leq k \leq \mathcal{U}} \in \mathcal{C}^{N_{t} \times \mathcal{U}}$ where $\boldsymbol{\hat{G}}_{l} = \boldsymbol{G}_{l} ( {\small \textbf{I}}_{N_{t}} - \boldsymbol{G}^{\dag}_{l,l} ( \boldsymbol{G}_{l,l} \boldsymbol{G}^{\dag}_{l,l} )^{-1} \boldsymbol{G}_{l,l} )$.\\
\textit{Proof:} See Appendix \ref{app:ZF-SIN}.
\end{prp}

Note that the proposed interference nulling-based precoder differs from the extended ZF scheme in \cite{6810573} where `all-zero' streams are sent for suppressing SI, i.e., in \cite{6810573} the (normalized) transmit signal vector $\mathbf{V}_{l} \left[ \begin{smallmatrix} \boldsymbol{s}_{l} \\ \mathbf{0}_{N_{r} \times 1} \end{smallmatrix} \right]$ with $\mathbf{V}_{l} \in \mathcal{C}^{N_{t} \times ( \mathcal{U} + N_{r} )}$ is set equal to the normalized columns of $\boldsymbol{\hat{G}}^{+}_{l} = \boldsymbol{\hat{G}}^{\dag}_{l} ( \boldsymbol{\hat{G}}_{l} \boldsymbol{\hat{G}}^{\dag}_{l} )^{-1} = [\boldsymbol{\hat{g}}_{l,k}]_{1 \leq j \leq \mathcal{U} + N_{r}}$ $\in \mathcal{C}^{N_{t} \times ( \mathcal{U} + N_{r} )}$ where $\boldsymbol{\hat{G}}_{l} = \left[ \begin{smallmatrix} \boldsymbol{G}_{l} \\ \mathbf{G}_{l,l} \end{smallmatrix} \right]$.

\begin{prp}
In the UL, a linear ZF decoder, eliminating multi-user interference, is employed with the normalized rows of $\boldsymbol{H}^{+}_{l} = ( \boldsymbol{H}_{l}^{\dag} \boldsymbol{H}_{l} )^{-1} \boldsymbol{H}^{\dag}_{l} = [\boldsymbol{\hat{h}}^{T}_{k,l}]^{T}_{1 \leq k \leq \mathcal{U}} \in \mathcal{C}^{\mathcal{U} \times N_{r}}$ set as the row vectors of $\boldsymbol{W}_{l}$, at the BS $l$.
\end{prp}

The received signal-to-interference-plus-noise ratio (SINR) in the DL is given by
\begin{align}
\gamma_{d} = \frac{\mathcal{X}_{d}}{\mathcal{I}_{d ,d} + \mathcal{I}_{u,d} + \mathcal{I}_{si,d} + \sigma^{2}_{d}}
\end{align}
where $\mathcal{X}_{d} = \tfrac{p^{(d)}}{\mathcal{U}} d_{b,o}^{- \beta} G_{b,o}$, $\mathcal{I}_{d,d} = \tfrac{p^{(d)}}{\mathcal{U}} \sum_{l \in \Phi^{(d)} \setminus \{ b \} } d_{l,o}^{- \beta} G_{l,o}$, $\mathcal{I}_{u,d} = \sum_{k \in \Phi^{(u)}_{l},l \in \Phi^{(d)} \setminus \{ o,b \} } p^{(u)}_{k,l} d_{k,o}^{- \beta} H_{k,o}$, $\mathcal{I}_{si,d} = p^{(u)}_{o,b} H_{o,o}$, $G_{b,o} \triangleq | \boldsymbol{g}_{b,o} \boldsymbol{v}_{b,o} |^2$, $G_{l,o} \triangleq \| \boldsymbol{g}_{l,o} \boldsymbol{V}_{l} \|^2$, $H_{k,o} \triangleq | h_{k,o} |^2$, and $H_{o,o} \triangleq | h_{o,o} |^2$. The ZF-SIN precoding vector $\boldsymbol{v}_{b,o} = \frac{\boldsymbol{\hat{g}}_{b,o}}{\| \boldsymbol{\hat{g}}_{b,o} \|}$ is selected in the direction of the projection of $\boldsymbol{g}_{b,0}$ on $\text{Null} ( [\boldsymbol{g}_{b,k}]_{1 \leq k \leq \mathcal{U}, k \neq 0},\boldsymbol{G}_{b,b} )$. The nullspace spanned by the SI and multi-user interference is $D_{d} \triangleq N_{t} - N_{r} - \mathcal{U} + 1$ dimensional. For the sake of analytical tractability, we assume that the outer-cell precoding matrices have independent column vectors \cite{7448861,7478073}. As a result, the channel power gain from each interfering BS in the DL is interpreted as the aggregation of multiple separate beams from the projection of the cross-link channel vector $\boldsymbol{g}_{l,o}$ onto the one-dimensional precoding vectors $\boldsymbol{v}_{l,k}$. The scheduled MTs, on the other hand, transmit using single-antennas (in all directions).

\begin{asm}
\label{asm:DL_Signals_Rayleigh}
The channel power gains at the reference MT $o$, from the intended BS $b$, interfering BS $l$, and scheduled MT $k$, are respectively $G_{b,o} \sim \Gamma (D_{d},1)$, $G_{l,o} \approx \Gamma (\mathcal{U},1)$, and $H_{k,o} \sim \Gamma (1,1)$. 
\end{asm}

\begin{rem}
Utilizing the linear ZF-SIN precoder for spatially suppressing SI at the BS side results in a loss of $N_{r}$ (number of receive antennas) DoF in the DL antenna array gain. 
\end{rem}

The Rayleigh fading model applies to cases without line-of-sight (LOS), e.g., with afar transceiver distances. In FD setups, however, the nodes' transmit and receive antennas are co-located. Hence, the Rician fading model, which takes into account the different LOS and scattered paths, can be invoked to capture performance under generalized SI cancellation capability \nolinebreak[4] \cite{7041186}.

\begin{asm}
\label{asm:UL_SI_Signal_Rician}

The SI channel power gain at the reference MT $o$ is a non-central Chi-squared random variable with Rician factor $K$ and fading attenuation $\Omega$ such that $\mu \triangleq \sqrt{\frac{K \Omega}{K + 1}}$ and $\nu \triangleq \sqrt{\frac{\Omega}{K+1}}$. The corresponding p.d.f. and m.g.f. are respectively given by  
\begin{align}
\mathcal{P}_{H_{o,o}} (h) & = \frac{1 + K}{\Omega} \exp \left( - \left( K + \frac{(1 + K) h}{\Omega} \right) \right) I_0\left(2 \sqrt{ \frac{K (1 + K) h}{\Omega} }\right) 
\label{eq:pdfChi}
\end{align}
and
\begin{align}
\mathcal{M}_{H_{o,o}} (z) = \frac{1 + K}{1 + K + z \Omega} \exp \left( - \frac{z K \Omega}{1 + K + z \Omega} \right).
\label{eq:mgfChi}
\end{align}
\end{asm}

\begin{rem}
The SI channel power gain at the reference MT $o$ can be approximated using Gamma moment matching as $H_{o,o} \sim \Gamma \left( \kappa,\theta \right)$ where $\kappa \triangleq \tfrac{ \big( \mu^2+\nu^2 \big)^2}{\left( 2 \mu^2 + \nu^2 \right) \nu^2 }$ and $\theta \triangleq \tfrac{\big( 2 \mu^2 + \nu^2 \big) \nu^2 }{\mu^2+\nu^2}$ \cite{7805138}.
\end{rem}

Next, we express the received SINR from the reference MT at its serving BS as
\begin{align}
\gamma_{u} = \frac{\mathcal{X}_{u}}{\mathcal{I}_{u,u} + \mathcal{I}_{d,u} + \sigma^{2}_{u}}
\end{align}
where $\mathcal{X}_{u} = p^{(u)}_{o,b} d_{o,b}^{- \beta} H_{o,b}$, $\mathcal{I}_{u,u} = \sum_{k \in \Phi^{(u)}_{l} , l \in \Phi^{(d)} \setminus \{ b \} } p^{(u)}_{k,l} d_{k,b}^{- \beta} H_{k,b}$, $\mathcal{I}_{d,u} = \tfrac{p^{(d)}}{\mathcal{U}} \sum_{l \in \Phi^{(d)} \setminus \{ b \} } d_{l,b}^{- \beta} G_{l,b}$, $H_{o,b} \triangleq | \boldsymbol{w}^{T}_{o,b} \boldsymbol{h}_{o,b} |^2$, $H_{k,b} \triangleq | \boldsymbol{w}^{T}_{o,b}\boldsymbol{h}_{k,b} |^2$, and $G_{l,b} \triangleq \| \boldsymbol{w}^{T}_{o,b} \boldsymbol{G}_{l,b} \boldsymbol{V}_{l} \|^2$. Note $\| \hat{\boldsymbol{h}}^{T}_{o,b} \|^{- 2} = \{ ( \boldsymbol{H}^{\dag}_{b} \boldsymbol{H}_{b})^{-1} \}_{o,o} \sim \text{Erlang} (D_{u},1)$ where $D_{u} \triangleq N_{r} - \mathcal{U} + 1$. In turn, $\boldsymbol{W}_{b}$ is selected independently from $\boldsymbol{h}_{k,b}$ and $\boldsymbol{G}_{l,b}$. We recall the assumption that $\boldsymbol{V}_{l}$ has independent column vectors.

\begin{asm}
\label{asm:UL_Signals_Rayleigh}
The channel power gains at the reference BS $b$, from the intended MT $o$, interfering MT $k$, and interfering BS $l$ are respectively modeled using $H_{o,b} \sim \Gamma \left( D_{u},1 \right)$, $H_{k,b} \sim \Gamma \left( 1 , 1 \right)$, and $G_{l,b} \approx \Gamma (\mathcal{U},1)$.    
\end{asm}

It is important to note that it is feasible to apply other linear precoding schemes such as CB in FD massive MIMO systems \cite{BaiS16}. In such cases, at the reference BS $b$, the SI channel power gain $G_{b,b} \triangleq \| \boldsymbol{w}^{T}_{o,b} \boldsymbol{G}_{b,b} \boldsymbol{V}_{b} \|^2$ needs to be characterized. In \cite{7805138}, the distribution of the SI with linear processing over FD multi-user MIMO Rician fading channels was derived using Gamma moment matching. In particular, for FD multi-user massive MIMO systems, the following holds. 

\begin{rem}
With conventional linear precoders in the DL (such as CB and ZF), the SI channel power gain at the reference massive MIMO BS $b$ can be approximated using Gamma moment matching as $G_{b,b} \sim \Gamma ( \kappa , \theta)$ where $\kappa = \tfrac{\mathcal{U} \big( \mu^2 + \nu^2 \big)^2}{(\mathcal{U}+2) \mu^4 + 2 \mu^2 \nu^2 + \nu^4}$ and $\theta = \tfrac{(\mathcal{U}+2) \mu^4 + 2 \mu^2 \nu^2 + \nu^4}{\mu^2 + \nu^2}$ \cite{7805138}. 
\end{rem}

\section{Self-Interference-Aware Power Control}

In long-term-evolution (LTE), UL fractional power control is defined to account for the path-loss effect \cite{5706315}. Recently, interference-aware fractional power control has been proposed to ensure that the power adjustment intended for path-loss compensation does not cause undesired interference to neighboring nodes \cite{MDR-IntAware-arXiv}. In this work, we propose an LTE-compliant SIA fractional power control mechanism where the MTs adjust their transmit power based on the distance-dependent path-loss, SI, and maximum available transmit power. Here, specifically, an arbitrary scheduled MT $k$ transmits to its serving BS $l$ using 
\begin{align}
p^{(u)}_{k,l} = \min \left(p_{0} d_{k,l}^{ \psi \beta},I_{\text{SI}} H^{-1}_{k,k},p^{(u)} \right)
\end{align}
where $p_{0}$, $\psi$ ($\in ( 0,1] $), $I_{\text{SI}}$, and $p^{(u)}$ are respectively the normalized power density, compensation factor, tolerable SI level, and maximum transmit power at the MT \cite{7448861}. The value of $I_{\text{SI}}$ can be set as the difference in the noise floor power from the gain of the MT SI cancellation capability.

\begin{lem}
\label{lem:Power_Distribution_SIA_Rician}
The c.d.f. and p.d.f. of the transmit power of a typical MT under the SIA fractional power control mechanism are respectively given by
\begin{align}
\!\! \mathcal{F}_{p^{(u)}_{k,l}} \left( p \right) = \left( 1 - \exp \left( - \Xi_{\ROME{1}} (p) \right) \left( 1 - Q_{1} \left(\sqrt{2 K}, \sqrt{2 (1+K) \Xi_{\ROME{2}} (p) } \right) \right) \right) \left( 1 - \mathcal{H} \left( p - p^{(u)} \right) \right) + \mathcal{H} \left( p - p^{(u)} \right) \!\!
\end{align}
and
\begin{equation}
\mathcal{P}_{p^{(u)}_{k,l}} \left( p \right) = \left\{ \,
\begin{IEEEeqnarraybox}[][c]{l?s}
\IEEEstrut
\exp \left( - \Xi_{\ROME{1}} \left( p^{(u)} \right) \right) \left( 1 - Q_{1} \left(\sqrt{2 K}, \sqrt{2 (1+K) \Xi_{\ROME{2}} \left( p^{(u)} \right) } \right) \right) \delta \left( p - p^{(u)} \right) & $p \geq p^{(u)}$ \\
\frac{\exp \left(- \Xi_{\ROME{1}} (p) \right)}{p} \Biggr[ \frac{2 \Xi_{\ROME{1}}(p)}{\psi \beta} \left( 1 - Q_{1} \left(\sqrt{2 K}, \sqrt{2 (1+K) \Xi_{\ROME{2}} (p) } \right) \right) + (1+K) \Xi_{\ROME{2}} (p) \\ \times \exp \left( - \left( K + (1+K) \Xi_{\ROME{2}} (p) \right)\right) \, _0\tilde{F}_1\left(;1;K (1+K) \Xi_{\ROME{2}} (p) \right) \Biggr] & $p < p^{(u)}$
\IEEEstrut
\end{IEEEeqnarraybox}
\right.
\label{eq:Power_PDF_SIA_Rician}
\end{equation} 
where $\Xi_{\ROME{1}}(p) = \pi \lambda^{(d)} \left( \frac{p}{p_{0}} \right)^{\frac{2}{\psi \beta}}$
and $\Xi_{\ROME{2}}(p) = \frac{I_{\text{SI}}}{p \Omega}$.\\
\textit{Proof:} See Appendix \ref{app:Power_Distribution_SIA_Rician}.
\end{lem}

\begin{rem}
The proposed SIA fractional power control mechanism is a generalization of the existing approaches for UL power control including total (without $I_{\text{SI}}$) and truncated (without $I_{\text{SI}}$ and $p^{(u)}$) fractional power control schemes. 
\end{rem}

The computation of SE can be greatly simplified with a non-direct methodology requiring only the moments of the random variables involved \cite{7277029}. Next, we develop results for the moments of the SIA power control in certain special cases. It should be noted that a Meijer-G function can be readily calculated using common software for numerical computation. A Meijer-G function can also be expressed in terms of a hypergeometric function based on the results from \cite{adamchik}. 

\begin{cor}
\label{cor:Power_PDF_SIA_SC}
The p.d.f. of the transmit power of a typical MT under the SIA fractional power control mechanism can be simplified in certain special cases. 

For $K = 0$ (Rayleigh SI channel),
\begin{equation}
\mathcal{P}_{p^{(u)}_{k,l}} \left( p \right) = \left\{ \,
\begin{IEEEeqnarraybox}[][c]{l?s}
\IEEEstrut
\exp \left( - \Xi_{\ROME{1}} \left( {p^{(u)}} \right) \right) \left( 1 - \exp \left( - \Xi_{\ROME{2}} \left( p^{(u)} \right) \right) \right) \delta \left( p - p^{(u)} \right) & $p \geq p^{(u)}$ \\
\frac{\exp \left(- \Xi_{\ROME{1}} (p) \right)}{p} \Biggr( \frac{2 \Xi_{\ROME{1}} (p)}{\psi \beta} \left( 1 - \exp \left( - \Xi_{\ROME{2}} (p) \right) \right) + \Xi_{\ROME{2}} (p) \exp \left( - \Xi_{\ROME{2}} (p) \right) \Biggr) & $p < p^{(u)}.$
\IEEEstrut
\end{IEEEeqnarraybox}
\right.
\label{eq:Power_PDF_SIA_Rayleigh}
\end{equation} 

For $p_{0} \rightarrow + \infty$ (no path-loss compensation), 
\begin{equation}
\mathcal{P}_{p^{(u)}_{k,l}} \left( p \right) = \left\{ \,
\begin{IEEEeqnarraybox}[][c]{l?s}
\IEEEstrut
\left(1-Q_1\left(\sqrt{2 K},\sqrt{2 (1+K) \Xi_{\ROME{2}} \left( p^{(u)} \right)} \right)\right) \delta \left( p-p^{(u)} \right) & $p \geq p^{(u)}$ \\ (1+K) \Xi_{\ROME{2}} \left( p^{2} \right) \exp \left( - \left( K + (1+K) \Xi_{\ROME{2}} (p) \right) \right) \, _0\tilde{F}_1\left(;1;K (1+K) \Xi_{\ROME{2}} (p) \right) & $p < p^{(u)}.$
\IEEEstrut
\end{IEEEeqnarraybox}
\right.
\label{eq:Power_PDF_SIA_Rician_woCompensation}
\end{equation} 

For $I_{\text{SI}} \rightarrow + \infty$ (no constraint on the SI),
\begin{equation}
\mathcal{P}_{p^{(u)}_{k,l}} \left( p \right) = \left\{ \,
\begin{IEEEeqnarraybox}[][c]{l?s}
\IEEEstrut
\left( 1 - \exp \left( - \Xi_{\ROME{1}} \left( p^{(u)} \right) \right) \right) \delta \left( p - p^{(u)} \right) & $p \geq p^{(u)}$ \\
\frac{2 \Xi_{\ROME{1}}(p)}{\psi \beta p} \exp \left(- \Xi_{\ROME{1}} (p) \right) & $p < p^{(u)}$.
\IEEEstrut
\end{IEEEeqnarraybox}
\right.
\label{eq:Power_PDF_SIA_Rician_woSIA}
\end{equation} 
\end{cor}

\begin{lem}
\label{lem:Power_Moments_SIA_SC} 
The $\flat$-th positive moment of the transmit power of a typical MT under the SIA fractional power control mechanism admits a closed-form expression in certain special cases. 

For $p^{(u)} \rightarrow + \infty$ (no constraint on the maximum transmit power), $K = 0$ (Rayleigh SI channel), $\psi = 1$ (compensation factor), and $\beta = 4$ (path-loss exponent),
\begin{gather}
\mathbb{E} \left\{ {p^{(u)}_{k,l}}^{\flat} \right\} = \frac{\hat{\Xi}^{\flat}_{\ROME{2}}}{\sqrt{\pi}} \Biggr( \MeijerG[\Bigg]{3}{0}{0}{3}{-\flat + 1,0,\frac{1}{2}}{}{\frac{\hat{\Xi}_{\ROME{1}}^2 \hat{\Xi}_{\ROME{2}}}{4}} - \frac{\hat{\Xi}_{\ROME{1}} \sqrt{\hat{\Xi}_{\ROME{2}}} }{2} \MeijerG[\Bigg]{3}{0}{0}{3}{- \flat - \frac{1}{2},0,\frac{1}{2}}{}{\frac{\hat{\Xi}_{\ROME{1}}^2 \hat{\Xi}_{\ROME{2}}}{4}} \Biggr) + \frac{(2 \flat)!}{\hat{\Xi}^{2 \flat}_{\ROME{1}}}
\label{eq:Power_Moments_SIA_Rayleigh_PL4}
\end{gather}
where $\hat{\Xi}_{\ROME{1}} = \frac{\pi \lambda^{(d)}}{p^{\frac{2}{\beta \psi}}_{0}}$ (here $\hat{\Xi}_{\ROME{1}} = \frac{\pi \lambda^{(d)}}{\sqrt{p_{0}}}$) and $\hat{\Xi}_{\ROME{2}} = \frac{I_{\text{SI}}}{\Omega}$. 
 
For $p_{0} \rightarrow + \infty$ (no path-loss compensation) and $p^{(u)} \rightarrow + \infty$ (no constraint on the maximum transmit power),
\begin{align}
\mathbb{E} \left\{ {p^{(u)}_{k,l}}^{\flat} \right\} = (1 + K)^{\flat} \hat{\Xi}^{\flat}_{\ROME{2}} \Gamma (1-\flat) \, _1F_1(\flat;1;-K).
\label{eq:Power_Moments_SIA_Rician_woCompensation}
\end{align}
Further, for $K = 0$ (Rayleigh SI channel),
\begin{align}
\mathbb{E} \left\{ {p^{(u)}_{k,l}}^{\flat} \right\} = \hat{\Xi}^{\flat}_{\ROME{2}} \Gamma (1 - \flat). 
\end{align}
 
For $p^{(u)} \rightarrow + \infty)$ (no constraint on the maximum transmit power) and $I_{\text{SI}} \rightarrow + \infty$ (no constraint on the SI),
\begin{align}
\mathbb{E} \left\{ {p^{(u)}_{k,l}}^{\flat} \right\} = \frac{\Gamma \left(\frac{\psi \beta \flat }{2}+1\right)}{\hat{\Xi}^{\frac{\psi \beta \flat}{2}}_{\ROME{1}}}. 
\label{eq:Power_Moments_SIA_Rician_woSIA}
\end{align}
Further, for $\psi = 1$ (compensation factor), and $\beta = 4$ (path-loss exponent),
\begin{align}
\mathbb{E} \left\{ {p^{(u)}_{k,l}}^{\flat} \right\} = \frac{\Gamma \left( 1 + 2 \flat \right)}{\hat{\Xi}^{2 \flat}_{\ROME{1}}}. 
\end{align}
\textit{Proof:} See Appendix \ref{app:Power_Moments_SIA_SC}.
\end{lem}

\section{Spectral Efficiency Analysis} 

In order to facilitate performance analysis and optimization, we provide a framework for the computation of the DL and UL SEs in the FD massive MIMO cellular network. We utilize a m.g.f.-based methodology, which avoids the need for the direct computation of the SINR p.d.f. and only requires the m.g.f.s of the different signals involved \cite{5407601}, \cite{6516171}.

\subsection{Downlink}

We proceed by providing an explicit expression for the calculation of the SE in the DL.

\begin{thm}
\label{lem:SE}
The DL SE in the FD massive MIMO cellular network is given by
\begin{align}
\mathcal{S}_{d,f} = \mathbb{E} \left\{ \log_{2} \left( 1 + \gamma_{d} \right) \right\} & = \log_{2} (e) \int^{+ \infty}_{0} \int^{+ \infty}_{0} \mathcal{M}_{\mathcal{I}_{si,d} | p}(z) \mathcal{M}_{\mathcal{I}_{u,d} | p}(z) \int^{+ \infty}_{0} \left( 1 - \mathcal{M}_{\mathcal{X}_{d} | r}(z) \right) \nonumber \\ & \times \mathcal{M}_{\mathcal{I}_{d,d} | r}(z) \frac{\exp \left( - z \sigma^{2}_{d} \right)}{z} \mathcal{P}_{p^{(u)}_{k,l}} \left( p \right) \mathcal{P}_{d_{b,o}} \left( r \right) \diff z \diff r \diff p,
\label{eq:SE_DL}
\end{align}
\textit{Proof:} The result follows directly from \cite[Lemma 1]{5407601}.
\end{thm}

Next, we provide explicit expressions for the conditional m.g.f.s of the different DL signals. 

\begin{lem}
\label{lem:Signals_MGFs}
The conditional m.g.f.s of the different DL signals in the FD massive MIMO cellular network are given by
\begin{align}
\mathcal{M}_{\mathcal{X}_{d} | r}(z) = \left( 1 + z \tfrac{p^{(d)}}{\mathcal{U}} r^{- \beta} \right)^{- D_{d}},
\label{eq:d}
\end{align}
\begin{multline}
\mathcal{M}_{\mathcal{I}_{d,d} | r}(z) = \exp \mathlarger{\Biggr(} - \pi  \lambda^{(d)} \mathlarger{\Biggr[} r^{2} \left(\left(z \tfrac{p^{(d)}}{\mathcal{U}} r^{-\beta} +1\right)^{- \mathcal{U}}-1\right) + \Gamma \left(\mathcal{U}+\frac{2}{\beta} \right) {\left( z \tfrac{p^{(d)}}{\mathcal{U}} \right)^{- \mathcal{U}}} \\ \times \left( \left( z \tfrac{p^{(d)}}{\mathcal{U}} \right)^{\mathcal{U} + \frac{2}{\beta }} \frac{\Gamma \left(1 - \frac{2}{\beta }\right)}{\Gamma \left( \mathcal{U} \right)} - \mathcal{U} r^{ \mathcal{U} \beta +2} \, _2\tilde{F}_1\left(\mathcal{U}+1,\mathcal{U}+\frac{2}{\beta};\mathcal{U}+\frac{2}{\beta }+1;- \frac{r^{\beta}}{z \tfrac{p^{(d)}}{\mathcal{U}}}\right) \right) \mathlarger{\Biggr]} \mathlarger{\Biggr)},
\label{eq:d->d}
\end{multline}
\begin{gather}
\mathcal{M}_{\mathcal{I}_{u,d} | p}(z) = \exp \left( - \pi \mathcal{U} \lambda^{(d)} \left( z p \right)^{\frac{2}{\beta}} \Gamma \left( 1 - \frac{2}{\beta} \right) \Gamma \left(1+\frac{2}{\beta }\right) \right),
\label{eq:u->d}
\end{gather}
and
\begin{align}
\mathcal{M}_{\mathcal{I}_{si,d} | p}(z) = \frac{1 + K}{1 + K + z p \Omega} \exp \left( - \frac{z p K \Omega}{1 + K + z p \Omega} \right).
\label{eq:si->d}
\end{align}
\textit{Proof:} See Appendix \ref{app:Signals_MGFs}.
\end{lem}

\begin{rem}
The CI at the reference MT in (\ref{eq:u->d}), is from the aggregation of interference from all other scheduled MTs; hence, it is approximated using an independent PPP with spatial density $\mathcal{U} \lambda^{(d)}$ based on \textit{Assumption \ref{asm:UEsProcess}}.
\end{rem}

\subsection{Uplink}

Here, we provide a systematic approach for the calculation of the SE in the UL.

\begin{thm}
\label{lem:SE}
The UL SE in the FD massive MIMO cellular network is given by
\begin{align}
\mathcal{S}_{u,f} = \mathbb{E} \left\{ \log_{2} \left( 1 + \gamma_{u} \right) \right\} & = \log_{2} (e) \int^{+ \infty}_{0} \mathcal{M}_{\mathcal{I}_{d,u}}(z) \int^{+ \infty}_{0} \int^{+ \infty}_{0} \left( 1 - \mathcal{M}_{\mathcal{X}_{u} | p,r}(z) \right) \mathcal{M}_{\mathcal{I}_{u,u} | p,r}(z) \nonumber \\ & \times \frac{\exp \left( - z \sigma^{2}_{u} \right)}{z} \mathcal{P}_{p^{(u)}_{k,l}} \left( p \right) \mathcal{P}_{d_{b,o}} \left( r \right) \diff z \diff r \diff p.
\label{eq:SE_UL}
\end{align}
\textit{Proof:} The result follows directly from \cite[Lemma 1]{5407601}.
\end{thm}

The conditional m.g.f.s of the signals required for the calculation of the UL SE can be derived in closed-form as in the following lemma. 

\begin{lem}
\label{lem:Signals_MGFs}
The conditional m.g.f.s of the different UL signals in the FD massive MIMO cellular network are given by
\begin{align}
\mathcal{M}_{\mathcal{X}_{u} | p,r}(z) & = \left( 1 + z p r^{- \beta} \right)^{- D_{u}},
\label{eq:u}
\end{align}
\begin{multline}
\mathcal{M}_{\mathcal{I}_{u,u} | p,r} ( z ) = \exp \Biggr( - \pi \mathcal{U} \lambda^{(d)} \Biggr[ \left( z p \right)^{\frac{2}{\beta}} \Gamma \left(1 - \frac{2}{\beta }\right) \Gamma \left(1+\frac{2}{\beta }\right) \\ - r^{2} \Biggr(1 - \frac{2 r^{\beta} }{ z p \left( \beta +2 \right) } \, _2F_1\left(1,1 + \frac{2}{\beta };2+\frac{2}{\beta };-\frac{ r^{\beta }}{z p} \right) \Biggr) \Biggr] \Biggr), 
\label{eq:u->u}
\end{multline}
and
\begin{gather}
\mathcal{M}_{\mathcal{I}_{d,u}} \left( z \right) = \exp \left( - \pi \lambda^{(d)} \left( z \tfrac{p^{(d)}}{\mathcal{U}} \right)^{\frac{2}{\beta}} \frac{\Gamma \left( 1 - \frac{2}{\beta} \right) \Gamma \left( \mathcal{U} + \frac{2}{\beta} \right)}{\mathcal{U}^{\frac{2}{\beta}} \Gamma \left( \mathcal{U} \right)} \right).
\label{eq:d->u}
\end{gather}
\textit{Proof:} See Appendix \ref{app:Signals_MGFs}.
\end{lem}

\begin{rem}
{\color{black} The UL inter-cell interference in (\ref{eq:u->u}) is derived using a spatially-thinned PPP with spatial density $\mathcal{U} \lambda^{(d)}$ and circular exclusion region with radius $r$ based on the spatial constraints from \textit{Assumption \ref{asm:UEsProcess}} \cite{7511710,7448861}.}
\end{rem}

\subsection{Special Cases}

In certain special cases, the conditional m.g.f.s of the signals can be simplified further as shown in the following results.

\begin{cor}
{\color{black} The conditional m.g.f. of the inter-cell interference in the DL for $\mathcal{U} = 1$ (single-user), and $\beta = 4$ (path-loss exponent) is given by
\begin{gather}
\mathcal{M}_{\mathcal{I}_{d,d} | r} ( z ) = \exp \left( - \pi \lambda^{(d)} \sqrt{z p^{(d)}} \arccot \left( \frac{r^2}{\sqrt{z p^{(d)}}} \right) \right).  
\end{gather}}
\end{cor}

\begin{cor}
The conditional m.g.f. of the CI in the DL with $\beta = 4$ (path-loss exponent) is given by  
\begin{gather}
\mathcal{M}_{\mathcal{I}_{u,d} | p}(z) = \exp \left( - \frac{\pi^{2}}{2} \mathcal{U} \lambda^{(d)} \sqrt{z p} \right).
\end{gather}
The above can be further simplified under special cases of UL power control. 

For $p^{(u)} \rightarrow + \infty$ (no constraint on the maximum transmit power), $K = 0$ (Rayleigh SI channel), $\psi = 1$ (compensation factor), and $\beta = 4$ (path-loss exponent),
\begin{multline}
\mathcal{M}_{\mathcal{I}_{u,d}} (z) = \frac{1}{1 + \frac{\pi}{2} \sqrt{z p_0}} + \frac{1}{\sqrt{\pi}} \MeijerG[\Bigg]{3}{0}{0}{3}{0,\frac{1}{2},1}{}{ \hat{\Xi}_{\ROME{2}} \left( \frac{\mathcal{U} \hat{\Xi}_{\ROME{1}}}{2} \left( 1 + \frac{\pi}{2} \sqrt{z p_0} \right) \right)^2} \\ - \frac{\mathcal{U} \tilde{\Xi}_{\ROME{1}} \sqrt{\tilde{\Xi}_{\ROME{2}}}}{2 \sqrt{\pi}} \MeijerG[\Bigg]{3}{0}{0}{3}{-\frac{1}{2},0,\frac{1}{2}}{}{ \hat{\Xi}_{\ROME{2}} \left( \frac{\mathcal{U} \hat{\Xi}_{\ROME{1}}}{2} \left( 1 + \frac{\pi}{2} \sqrt{z p_0} \right) \right)^2}.
\label{eqn:tn}
\end{multline}
Further, for $p_{0} \rightarrow + \infty$ (no path-loss compensation),
\begin{align}
\mathcal{M}_{\mathcal{I}_{u,d}} (z) = \frac{1}{\sqrt{\pi}} \MeijerG[\Bigg]{3}{0}{0}{3}{0,\frac{1}{2},1}{}{ z \hat{\Xi}_{\ROME{2}} \left( \frac{\pi^{2} \mathcal{U} \lambda^{(d)}}{4} \right)^{2}}.
\end{align}
On the other hand, for $I_{\text{SI}} \rightarrow + \infty$ (no constraint on the SI), 
\begin{align}
\mathcal{M}_{\mathcal{I}_{u,d}} (z) = \frac{1}{1 + \frac{\pi}{2} \sqrt{z p_{0}}}.
\end{align}
\end{cor}

\begin{cor}
The conditional m.g.f. of the inter-cell interference in the UL for $\beta = 4$ (path-loss exponent) is given by
\begin{gather}
\mathcal{M}_{\mathcal{I}_{u,u} | p,r} (z) = \exp \left( - \pi \mathcal{U} \lambda^{(d)} \sqrt{z p} \arccot \left( \frac{r^2}{\sqrt{z p}} \right) \right).
\end{gather}
\end{cor}

\begin{cor}
The conditional m.g.f. of the CI in the UL
for $\mathcal{U} = 1$ (single-user) and $\beta = 4$ (path-loss exponent) is given by
\begin{gather}
\mathcal{M}_{\mathcal{I}_{d,u}} \left( z \right) = \exp \left( - \frac{\pi^{2}}{2} \lambda^{(d)} \sqrt{z p^{(d)}} \right).
\end{gather}
\end{cor}

The SE expressions in (\ref{eq:SE_DL}) and (\ref{eq:SE_UL}) require three-fold integral computations - as opposed to the manifold integrals involved in the direct p.d.f.-based approach. It is possible to further reduce the computational complexity in certain special cases with UL power control. For baseline SISO, we utilize linear ZF precoding, and take into account the SI at the BS side in FD mode.

\begin{lem}
For $N_{t},N_{r},\mathcal{U} = 1$ (baseline SISO), $K = 0$ (Rayleigh SI channel), $\sigma^{2}_{d} , \sigma^{2}_{u} = 0$ (interference-limited region), and $\beta = 4$ (path-loss exponent), the SEs ($\omega = \frac{p^{(d)}}{p}$ for DL and $\omega = \frac{p}{p^{(d)}}$ for UL) can be reduced to double-integral expressions
\begin{multline}
\mathcal{S}_{\{.\},f} = \log_{2} (e) \int^{+ \infty}_{0} \int^{+ \infty}_{0} \frac{2 \pi \lambda^{(d)}}{\sqrt{\Omega} \left(1 + s^{2} \right)} \sqrt{\omega} \mathlarger{\Biggr[} \sin \left( \frac{\pi \lambda^{(d)}}{\sqrt{\Omega}} \left(\frac{\pi}{2} + \sqrt{\omega} \left(s+\arccot(s)\right) \right) \right) \\ \times \mathscr{C} \left( \frac{\pi \lambda^{(d)}}{\sqrt{\Omega}} \left(\frac{\pi}{2} + \sqrt{\omega} \left(s+\arccot(s)\right) \right) \right) + \cos \left( \frac{\pi \lambda^{(d)}}{\sqrt{\Omega}} \left(\frac{\pi}{2} + \sqrt{\omega} \left(s+\arccot(s)\right) \right) \right) \\ \times \left(\frac{\pi}{2}-\mathscr{S} \left( \frac{\pi \lambda^{(d)}}{\sqrt{\Omega}} \left(\frac{\pi}{2} + \sqrt{\omega} \left(s+\arccot(s)\right) \right) \right) \right) \mathlarger{\Biggr]} \mathcal{P}_{p^{(u)}_{k,l}} \left( p \right) \diff s \diff p.
\label{eq:SE_SC}
\end{multline}
Further, for $\text{SI} = 0$ (perfect SI subtraction), the SEs ($\omega = \frac{p^{(d)}}{p}$ for DL and $\omega = \frac{p}{p^{(d)}}$ for UL) are given by
\begin{align}
\mathcal{S}_{\{.\},f}  = \log_{2} (e) \int^{+ \infty}_{0} \int^{+ \infty}_{0} \frac{2}{\left( 1 + s^2 \right) \left(\frac{\pi}{2 \sqrt{\omega}} + s + \arccot (s) \right)} \mathcal{P}_{p^{(u)}_{k,l}} \left( p \right) \diff s \diff p.
\label{eq:SE_SC_noSI}
\end{align}

For $p_{0} \rightarrow + \infty$ (no path-loss compensation) and $p^{(u)} \rightarrow + \infty$ (no constraint on the maximum transmit power), the SE expressions in (\ref{eq:SE_SC_noSI}) can be reduced to single-fold integrals
\begin{multline}
\mathcal{S}_{d,f} = \log_{2} (e) \int^{+ \infty}_{0} \frac{2}{\left(1 + s^2\right) \left(s+\arccot(s)\right)} \Biggr[ 1 - \frac{\pi \sqrt{\pi}}{2 \left(s + \arccot (s)\right)} \sqrt{\frac{I_{\text{SI}}}{p^{(d)} \Omega}} \\ - \left(\frac{\pi }{2 \left(s+ \arccot (s)\right)}\right)^2 \frac{I_{\text{SI}}}{p^{(d)} \Omega} \exp \left( - \left(\frac{\pi}{2 \left(s + \arccot (s) \right)}\right)^2 \frac{I_{\text{SI}}}{p^{(d)} \Omega} \right) \\ \times \Biggr(\text{Ei}\left( \left(\frac{\pi }{2 \left(s+ \arccot (s)\right)}\right)^2 \frac{I_{\text{SI}}}{p^{(d)} \Omega} \right) - \pi \text{erfi} \left( \frac{\pi }{2 \left(s+ \arccot (s)\right)} \sqrt{\frac{I_{\text{SI}}}{p^{(d)} \Omega}} \right)
 \Biggr) \Biggr] \diff s 
\label{eq:SE_DL_SC_woComp}
\end{multline}
and
\begin{multline}
\mathcal{S}_{u,f} = \log_{2} (e) \int^{+ \infty}_{0} \frac{4}{\pi \left( 1 + s^2 \right)} \sqrt{\frac{I_{\text{SI}}}{p^{(d)} \Omega}} \Biggr[ \sqrt{\pi} + \frac{2 \left( s + \arccot (s) \right)}{\pi} \sqrt{\frac{I_{\text{SI}}}{p^{(d)} \Omega}} \exp \Biggr( - \left( \frac{2 \left( s + \arccot (s) \right)}{\pi} \right)^2 \\ \times \frac{I_{\text{SI}}}{p^{(d)} \Omega} \Biggr) \Biggr( \text{Ei}\left( \left( \frac{2 \left( s + \arccot (s) \right)}{\pi} \right)^{2} \frac{I_{\text{SI}}}{p^{(d)} \Omega} \right) - \pi \text{erfi}\left( \frac{2 \left( s + \arccot (s) \right)}{\pi} \sqrt{\frac{I_{\text{SI}}}{p^{(d)} \Omega}} \right) \Biggr) \Biggr] \diff s. 
\label{eq:SE_UL_SC_woComp}
\end{multline}

For $I_{\text{SI}} \rightarrow + \infty$ (no constraint on the SI) and $p^{(u)} \rightarrow + \infty$ (no constraint on the maximum transmit power), the SE expressions in (\ref{eq:SE_SC_noSI}) can be reduced to single-fold integrals
\begin{gather}
\! \mathcal{S}_{d,f} = \log_{2} (e) \int^{+ \infty}_{0} \! \frac{- 4 \lambda^{(d)}}{1 + s^{2}} \sqrt{\frac{p^{(d)}}{p_{0}}} \exp \left( 2 \lambda^{(d)} \left(s + \arccot (s)\right) \sqrt{\frac{p^{(d)}}{p_{0}}} \right) \! \text{Ei}\left(-2 \lambda^{(d)} \left(s + \arccot (s) \right) \sqrt{\frac{p^{(d)}}{p_{0}}} \right) \diff s
\label{eq:SE_DL_SC_woSI}
\end{gather}
and
\begin{multline}
\mathcal{S}_{u,f} = \log_{2} (e) \int^{+ \infty}_{0} \frac{2}{\left( 1 + s^{2} \right) \left( s + \arccot (s) \right)} \mathlarger{\Biggr[} 1 + \frac{\pi^{2} \lambda^{(d)}}{2 \left( s + \arccot (s) \right)} \sqrt{\frac{p^{(d)}}{p_{0}}} \\ \times \exp \left( \frac{\pi^{2} \lambda^{(d)}}{2 \left( s + \arccot (s) \right)} \sqrt{\frac{p^{(d)}}{p_{0}}} \right) \text{Ei} \left( - \frac{\pi^{2} \lambda^{(d)}}{2 \left( s + \arccot (s) \right)} \sqrt{\frac{p^{(d)}}{p_{0}}} \right) \mathlarger{\Biggr]} \diff s.
\label{eq:SE_UL_SC_woSI}
\end{multline}
\textit{Proof:} See Appendix \ref{app:SE}.
\end{lem}

\section{Full-Duplex versus Half-Duplex}

The SE expressions developed facilitate performance analysis and optimization for generalized FD cellular deployments. At the same time, the proposed framework can serve as a benchmark tool for comparing the performance of FD over HD systems. Although an explicit expression for the corresponding gain cannot be obtained due to the highly complex SE expressions involving multiple improper integrals, we do provide results in certain special cases. 

In what follows, $\mathcal{S}_{d,h}$ and $\mathcal{S}_{u,h}$ respectively denote
the SEs of a typical MT in the DL and the UL of a HD cellular system. For comparison of FD and HD systems, we consider the SE over two time slots, i.e., $\mathbbm{S}_{f} = 2 (\mathcal{S}_{d,f} + \mathcal{S}_{u,f} )$ for FD, and $\mathbbm{S}_{h} = \mathcal{S}_{d,h} + \mathcal{S}_{u,h}$ for HD, respectively. We first study the baseline SISO case and then derive results for multi-user massive MIMO setups.

\subsection{Baseline SISO}

Here, we compare the FD versus HD performance for the baseline SISO case. 

\begin{lem}
\label{lem:FDvsHD}
The FD and HD SEs for $N^{t},N^{r},\mathcal{U} = 1$ (baseline SISO), $p^{(u)}$ (fixed MT transmit power), $\text{SI} = 0$ (perfect SI subtraction), $\sigma^{2}_{d} , \sigma^{2}_{u} = 0$ (interference-limited region), and $\beta = 4$ (path-loss exponent) are respectively given by
\begin{align}
\mathbbm{S}_{f} = \log_{2} (e) \! \int^{+ \infty}_{0} \! \frac{\frac{2 \pi}{1 + s^2} \left( \sqrt{\frac{p^{(u)}}{p^{(d)}}} + \sqrt{\frac{p^{(d)}}{p^{(u)}}} \right) + 8 ( s + \arccot (s) )}{ \left(\! \frac{\pi}{2} \sqrt{\frac{p^{(u)}}{p^{(d)}}} + s + \arccot (s) \! \right) \!\! \left(\! \frac{\pi}{2} \sqrt{\frac{p^{(d)}}{p^{(u)}}} + s + \arccot (s) \! \right)} \diff s
\label{eq:FD_SE_SC_EXACT}
\end{align}
and
\begin{align}
\mathbbm{S}_{h} = \log_{2} (e) \int^{+ \infty}_{0} \frac{4}{\left( 1 + s^{2} \right) \left( s  + \arccot(s) \right)} \diff s.  
\label{eq:HD_SE_SC_EXACT}
\end{align}
Further, bounded closed-form expressions of the FD and HD SEs are respectively given by 
\begin{align}
\tilde{\mathbbm{S}}_{f} \leq 2 \log_{2} (e) \left( \Psi \left( \frac{8}{\pi \left( 1 + \sqrt{\frac{p^{(u)}}{p^{(d)}}} \right)} - 1 \right) + \Psi \left( \frac{8}{\pi \left( 1 + \sqrt{\frac{p^{(d)}}{p^{(u)}}} \right)} - 1 \right) \right), \;\: \left( \frac{\pi }{\pi -8} \right)^{2} < \frac{p^{(d)}}{p^{(u)}} < \left( \frac{\pi -8}{\pi} \right)^{2} 
\label{eq:FD_SE_APPROX}
\end{align}
and
\begin{align}
\tilde{\mathbbm{S}}_{h} \leq 2 \log_{2} (e) \Psi \left( \frac{8}{\pi} - 1 \right)
\label{eq:HD_SE_APPROX}
\end{align}
where
\begin{multline}
\Psi \left( \alpha \right) = \frac{\left( 1 + \alpha \right) \left( \left( 5 - \alpha \right) \arccot \left( \sqrt{\alpha} \right) + \sqrt{\alpha} \log \left( \frac{1}{4} \left( 1 + \alpha \right) \right)  \right) - \frac{\pi}{2} \left( 5 - 4 \sqrt{\alpha} + \alpha \right) \left( 1 - 2 \sqrt{\alpha} - \alpha \right)}{\frac{\sqrt{\alpha}}{1 + \alpha} \left(  25 - 6 \alpha + \alpha^2 \right)}.
\label{eq:SE_SC_APPROX_FUNC} 
\end{multline}
\textit{Proof:} See Appendix \ref{app:FDvsHD}.
\end{lem}

\begin{rem}
The HD SE for the special case in Lemma \ref{lem:FDvsHD} is independent of the system parameters. The exact and bounded HD baseline SISO SEs over two time slots (in nat/s/Hz) are approximately $3$ and $2.72$, respectively. On the other hand, the exact and bounded FD baseline SISO SEs for the special case in Lemma \ref{lem:FDvsHD} are functions of the BS and MT transmit powers only.
\end{rem}

Based on the above remark, we next study the problem of finding the optimal fixed transmit powers which result in the largest FD over HD SE gain for baseline SISO. The SE function in HD mode, $\mathbbm{S}_{h}$ in (\ref{eq:HD_SE_SC_EXACT}), is affine in $p^{(d)} > 0$ and $p^{(u)} > 0$. On the other hand, it can be shown that the SE function in FD mode, $\mathbbm{S}_{f}$ in (\ref{eq:FD_SE_SC_EXACT}), is strictly quasi-concave in $p^{(d)} > 0$ and $p^{(u)} > 0$. This guarantees the existence of a unique maximum solution under a positive transmit power region. $\mathbbm{S}_{h}$ and $\mathbbm{S}_{f}$, however, involve improper integrals. Thus, an exact closed-form solution cannot be obtained analytically. Hence, we utilize the bounded SE expressions developed in (\ref{eq:FD_SE_APPROX}) and (\ref{eq:HD_SE_APPROX}).

\begin{lem}
\label{lem:optX_FDvsHD}

Consider the bounded FD and HD SEs for $N^{t},N^{r},\mathcal{U} = 1$ (baseline SISO), $p^{(u)}$ (MT transmit power), $\text{SI} = 0$ (perfect SI subtraction), $\sigma^{2}_{d} , \sigma^{2}_{u} = 0$ (interference-limited region), and $\beta = 4$ (path-loss exponent) from Lemma \ref{lem:FDvsHD}. We can formulate the following optimization problem for the highest FD over HD SE gain
\begin{align}
& \underset{x = \frac{p^{(d)}}{p^{(u)}}}{\text{maximize}} \quad \frac{\mathbbm{S}_{f}}{\mathbbm{S}_{h}} = \frac{\Psi \left( \frac{8}{\pi \left( 1 + \sqrt{x} \right)}-1 \right) + \Psi \left( \frac{8}{\pi \left( 1 + \frac{1}{\sqrt{x}} \right)}-1 \right)}{\Psi \left( \frac{8}{\pi}-1 \right)} \quad \text{subject to:} \quad \left( \frac{\pi }{\pi -8} \right)^{2} < x < \left( \frac{\pi -8}{\pi} \right)^{2}.
\label{eq:opt_prob_power_simplified}
\end{align}
The solution to the above problem is given by 
\begin{align}
x^{*} = 1, \quad \left( \frac{\mathbbm{S}_{f}}{\mathbbm{S}_{h}} \right)^{*} = \frac{ 2 \Psi \left( \frac{4}{\pi}-1 \right)}{\Psi \left( \frac{8}{\pi}-1 \right)} = 1 + 13 \left( \frac{1}{10^{2}} \right) + \frac{1}{2} \left( \frac{1}{10^4} \right).
\end{align}
\textit{Proof:} See Appendix \ref{app:optX_FDvsHD}.
\end{lem}

\begin{rem}
From Lemma \ref{lem:optX_FDvsHD}, the highest exact and bounded SE percentage gains of FD over HD for baseline SISO are respectively $\sim$9\% and $\sim$13\%.
We can infer that without advanced techniques for tackling the CI, even with perfect SI subtraction,  the FD baseline SISO system achieves only modest improvements over its HD counterpart.     
\end{rem}

\subsection{Massive MIMO}

Next, we analyze the FD versus HD SE for massive MIMO-enabled cellular networks. 

\begin{lem}
\label{lem:FDvsHDmm}
The FD and HD SEs with massive MIMO, ZF beamforming, $\mathcal{U} p$ (BS transmit power), $p$ (MT transmit power), $\text{SI} = 0$ (perfect SI subtraction), $\sigma^{2}_{d} , \sigma^{2}_{u} = 0$ (interference-limited region), and $\beta = 4$ (path-loss exponent) are respectively given by
\begin{multline}
\mathbbm{S}_{f} = \log_{2} (e) \int^{+ \infty}_{0} \frac{4}{1 + s^{2}} \mathlarger{\Biggr(} \frac{ 1 + s^2 \left(1-\left( 1 + \frac{1}{s^2} \right)^{\mathcal{U}-N_{r}}\right) }{ s + \mathcal{U} \arccot (s) + \sqrt{\frac{\pi}{\mathcal{U}}} \frac{\Gamma \left(\mathcal{U}+\frac{1}{2}\right)}{\Gamma (\mathcal{U})} } \\ + \frac{ 1 + s^2 \left( 1 - \left( 1 + \frac{1}{s^2} \right)^{\mathcal{U}-N_{t}} \right) }{ \frac{s}{\mathcal{U}} \left(1 + \frac{1}{s^2} \right)^{-\mathcal{U}} - s^{2 \mathcal{U}+1} \Gamma \left(\mathcal{U}+\frac{1}{2}\right) \, _2\tilde{F}_1\left(\mathcal{U}+\frac{1}{2},\mathcal{U}+1;\mathcal{U}+\frac{3}{2};-s^2\right) + \sqrt{\pi } \frac{\Gamma \left(\mathcal{U}+\frac{1}{2}\right)}{ \mathcal{U} \Gamma (\mathcal{U}) } + \frac{\pi}{2} } \mathlarger{\Biggr)} \diff s
\label{eq:FD_SE_SC_EXACTmm}
\end{multline}
and
\begin{multline}
\mathbbm{S}_{h} = \log_{2} (e) \int^{+ \infty}_{0} \frac{2}{1 + s^2} \mathlarger{\Biggr(} \frac{ 1 + s^2 \left(1 - \left( 1 + \frac{1}{s^2} \right)^{\mathcal{U}-N_{r}}\right) }{ s + \mathcal{U} \arccot (s) } \\ + \frac{ 1 + s^2 \left( 1 - \left( 1 + \frac{1}{s^2} \right)^{\mathcal{U} - N_{t}} \right)}{ s \left(1 + \frac{1}{s^2} \right)^{-\mathcal{U}} - \mathcal{U} s^{2 \mathcal{U} + 1} \Gamma \left(\mathcal{U}+\frac{1}{2}\right) \, _2\tilde{F}_1\left(\mathcal{U}+\frac{1}{2},\mathcal{U}+1;\mathcal{U}+\frac{3}{2};-s^2\right) + \sqrt{\pi } \frac{\Gamma \left(\mathcal{U}+\frac{1}{2}\right)}{\Gamma (\mathcal{U}) } } \mathlarger{\Biggr)} \diff s. 
\label{eq:HD_SE_SC_EXACTmm}
\end{multline}
For $\mathcal{U} = 1$ (single-user), the above expressions can be respectively reduced to
\begin{align}
\mathbbm{S}_{f} = \log_{2} (e) \int^{+ \infty}_{0} \frac{4}{ s + \arccot (s) + \frac{\pi}{2}} \left(2 - \left( 1 + \frac{1}{s^2} \right)^{- N_{r}}- \left( 1 + \frac{1}{s^2} \right)^{- N_{t}} \right) \diff s
\end{align}
and
\begin{align}
\mathbbm{S}_{h} = \log_{2} (e) \int^{+ \infty}_{0} \frac{2}{s + \arccot(s) } \left(2 - \left( 1 + \frac{1}{s^2} \right)^{- N_{r}} - \left( 1 + \frac{1}{s^2} \right)^{- N_{t}} \right) \diff s.
\end{align}
\textit{Proof:} The proof follows from a similar approach to that in Appendix \ref{app:FDvsHD}.
\end{lem}

\begin{rem}
By direct inspection of Lemma \ref{lem:FDvsHDmm}, we observe that the SEs in massive MIMO-enabled FD and HD cellular networks are functions of the number of antennas and users. 
\end{rem}

The SE expressions for the case of multi-user massive MIMO in Lemma \ref{lem:FDvsHDmm} involve single-fold integrals. Next, we employ non-linear curve-fitting in order to provide a closed-form approximation for the FD versus HD massive MIMO SE gain. 

\begin{cor}
\label{cor:NLCF}
The FD over HD massive MIMO SE gain with ZF beamforming, $\mathcal{N}$ (number of transmit and receive antennas), $\mathcal{U} p$ (BS transmit power), $p$ (MT transmit power), $\text{SI} = 0$ (perfect SI subtraction), $\sigma^{2}_{d} , \sigma^{2}_{u} = 0$ (interference-limited region), and $\beta = 4$ (path-loss exponent) can be approximated using non-linear curve fitting as\footnote{The properties related to the goodness of fit of the closed-form approximation in (\ref{eq:NLCFres}) are $R^{2} \approx 0.999$ and estimated variance of $\approx 1.17 \times 10^{- 3}$.}
\begin{gather}
\frac{\mathbbm{S}_{f}}{\mathbbm{S}_{h}} \approx 2 - \frac{9}{10} \left( \frac{\mathcal{U}^{\frac{4}{25}}}{\mathcal{N}^{\frac{1}{10}}} \right).
\label{eq:NLCFres}
\end{gather}
\end{cor}

\begin{rem}
We infer from the results of Lemma \ref{lem:FDvsHDmm} and Corollary \ref{cor:NLCF} that the FD over HD multi-user massive MIMO SE gain logarithmically increases (decreases) in the number of antennas (users). Furthermore, the anticipated two-fold increase in SE from massive MIMO-enabled FD versus HD cellular networks is achieved as the number of antennas tends to be infinitely large ($\mathcal{N} \rightarrow + \infty$).
\end{rem}

Note that EE can be studied (i) using the proposed framework for the calculation of SE and, (ii) adopting an appropriate power consumption model \cite{7031971}. A rigorous assessment of EE in FD massive MIMO cellular networks is left for future work. 

\section{Numerical Results}

In this section, we provide numerical examples to assess the performance of the FD massive MIMO cellular network under different system settings. MC simulations are accordingly provided for the purpose of examining the validity of the proposed analytical framework. The BS deployment density is considered to be $\lambda^{(d)} = \frac{4}{\pi}$ per unit area (km$\times$km). The total system bandwidth is $B = 20$ MHz. The noise power is calculated using $\sigma^{2} = -170 + 10 \log_{10} \left( B \right) + N_{f}$ (dBm), where $N_{f}$ is the noise figure \cite{7302534}. The maximum available transmit powers at the BSs and MTs are set to $43$ dBm and $23$ dBm, respectively. Note that the results from the MC simulations are obtained from $10^{4}$ trials in a circular region of radius $10^{2}$ km, considering (i) PPP under spatial constraints for the scheduled MTs (\textit{Assumption \ref{asm:UEsProcess}}), (ii) non-central Chi-squared distribution for the SI channel power gain (\textit{Assumption \ref{asm:UL_SI_Signal_Rician}}), and (iii) Gamma distribution for the other channels power gains (\textit{Assumptions \ref{asm:DL_Signals_Rayleigh} and \ref{asm:UL_Signals_Rayleigh}}).    

\begin{figure}[!t]
\centering
\includegraphics[scale=1]{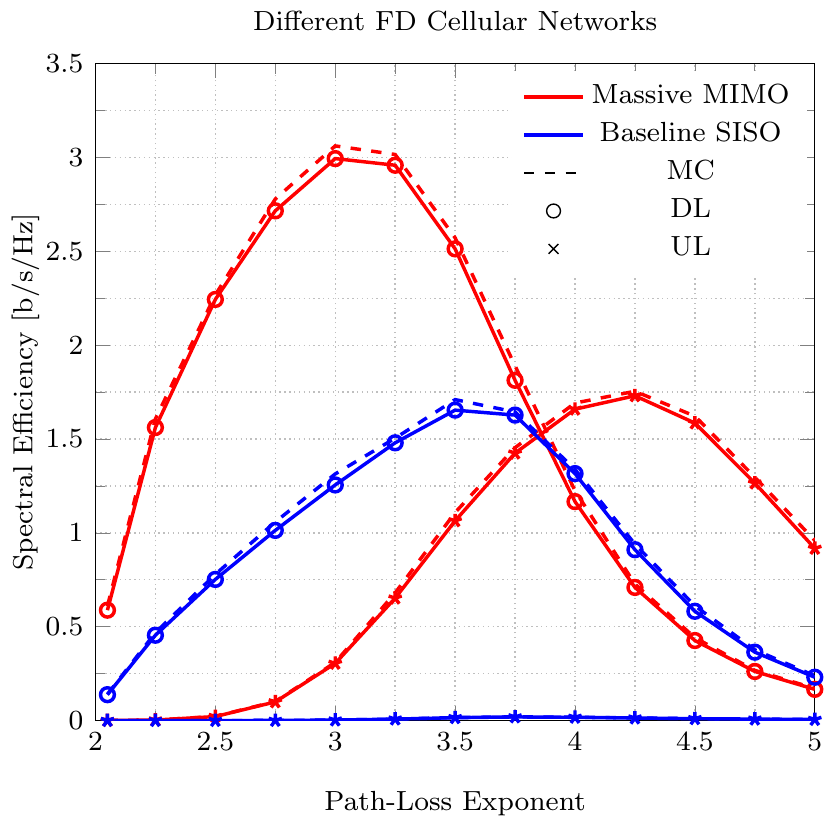}
\caption{System parameters are: $\lambda^{(d)} = \frac{4}{\pi}$ BSs/km$^{2}$, massive MIMO ($N_{t} = 80$, $N_{r} = 20$, $\mathcal{U} = 8$, $p^{(d)} = 30$ dBm), baseline SISO ($N_{t} = 1$, $N_{r} = 1$, $\mathcal{U} = 1$, $p^{(d)} = 43$ dBm), $p^{(u)} = 23$ dBm, $p_{0} = -80$ dBm, $B = 20$ MHz, $N_{f} = 10$ dB, $\psi = 1$, $\Omega = -80$ dB, $K = 1$.}
\label{SEvsSISIvsMIMOvsLSAS}
\end{figure}

\begin{figure}[!t]
\centering
\includegraphics[scale=1]{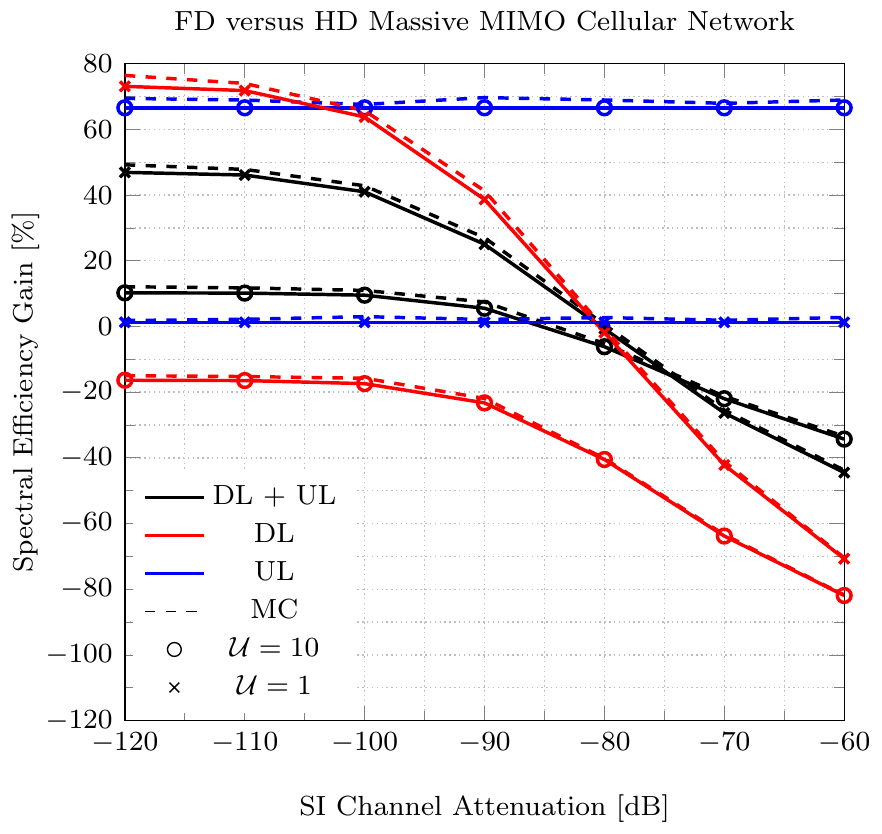}
\caption{System parameters are: $\lambda^{(d)} = \frac{4}{\pi}$ BSs/km$^{2}$, $N_{t} = 350$, $N_{r} = 50$, $p^{(d)} = 30$ dBm, $p^{(u)} = 23$ dBm, $p_{0} = -80$ dBm, $B = 20$ MHz, $N_{f} = 10$ dB, $\beta = 4$, $\psi = 1$, $K = 1$.}
\label{SEvsFDvsHD}
\end{figure}

\subsubsection{Different FD Cellular Setups} 

We compare the DL and UL performance of different FD cellular networks, namely, massive MIMO (with ZF-SIN precoder and ZF decoder) and baseline SISO (with ZF precoder/decoder) in Fig. \ref{SEvsSISIvsMIMOvsLSAS}. The results confirm prior findings that the UL rate is the main performance bottleneck in FD cellular systems with baseline SISO. Here, the UL performance is severely limited due to the imperfect SI subtraction and large disparity in the BS and MT transmit power levels. Furthermore, it can be observed that significant performance gains can be achieved by exploiting the large scale antenna array with linear ZF-SIN precoding and linear ZF receive combining. For example, with $\beta = 4$, the UL SE in the proposed FD massive MIMO cellular setup is more than $96$ times greater than that in the baseline SISO case. By increasing the antenna array size, further improvements can be realized from (i) the added DoF, and (ii) the potential to linearly reduce the transmit powers of the BSs and MTs without degrading the received signal-to-noise ratio (SNR). It should be noted that the MC results confirm the validity of the proposed analytical framework.

\begin{figure}[!t]
\centering
\includegraphics[scale=1]{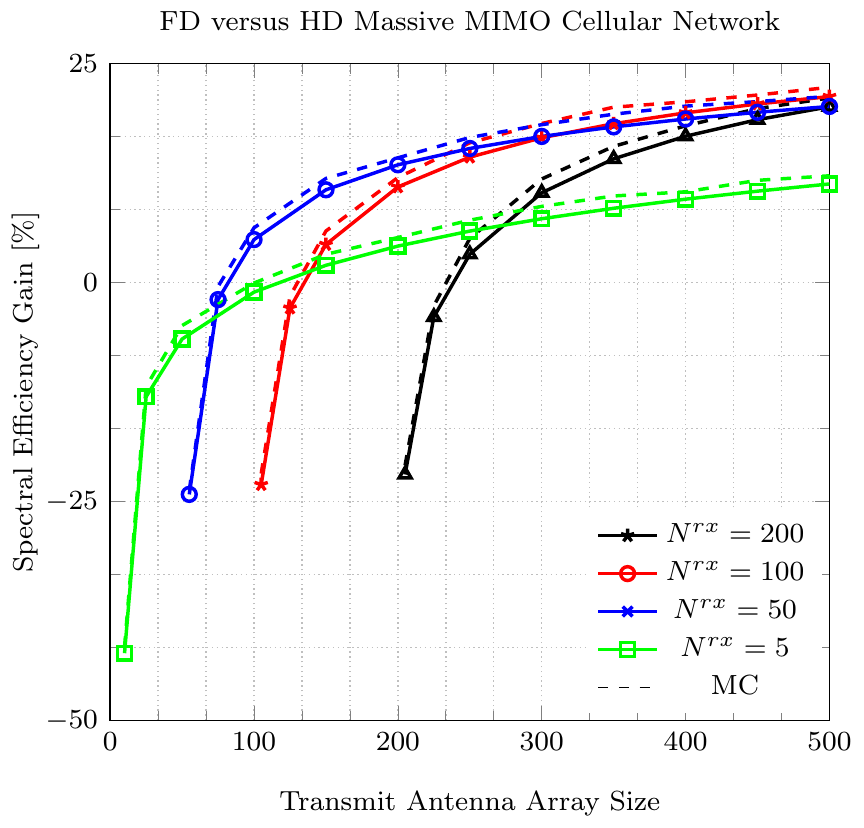}
\caption{System parameters are: $\lambda^{(d)} = \frac{4}{\pi}$ BSs/km$^{2}$, $\mathcal{U} = 5$, $p^{(d)} = 30$ dBm, $p^{(u)} = 23$ dBm, $p_{0} = -80$ dBm, $B = 20$ MHz, $N_{f} = 10$ dB, $\beta = 4$, $\psi = 1$, $\Omega = -90$ dB, $K = 2$.}
\label{SEvsAntennasvsFDvsHD}
\end{figure}

\subsubsection{FD versus HD Massive MIMO} Next, we investigate the performance of the FD massive MIMO cellular network with respect to its HD counterpart over a wide range of MT SI channel attenuations in Fig. \ref{SEvsFDvsHD}. It can be seen that any potential improvements from the FD operation occurs for SI channel attenuation well below $-80$ dB. This trend indicates that without advanced SI mitigation solutions being available at the MTs, the conventional HD massive MIMO cellular network is arguably the more sensible deployment choice. With nearly perfect SI cancellation at the reference MT, on the other hand, the maximum system SE gain in the FD massive MIMO cellular network over its analogous HD variant here is $47$\% with $\mathcal{U} = 1$ (resulting from $73$\% and $1$\% increase in the DL and UL SEs, respectively). It should be noted that the small improvement in UL SE is due to the CI from the BSs transmitting in the DL in FD mode.

\subsubsection{Transmit/Receive Antenna Array Size} The impact of different number of transmit and receive antennas on the SE gain of the massive MIMO-enabled FD system over its HD counterpart is depicted in Fig. \ref{SEvsAntennasvsFDvsHD}. It can be observed that the corresponding performance gain increases by adding more transmit antennas due to the reduced impact of the DL array gain from the ZF-SIN precoding scheme and improved resilience against interference. For smaller transmit antenna arrays, however, increasing the number of receive antennas may degrade the relative improvement from the FD operation over HD mode. It is also important to note that the relative FD over HD massive MIMO gain improves in the path-loss severity. The optimal ratio of transmit over receive antennas, on the other hand, depends on the particular system settings. A rigorous study of this aspect is postponed to future work. 

\begin{figure}[!t]
\centering
\includegraphics[scale=1]{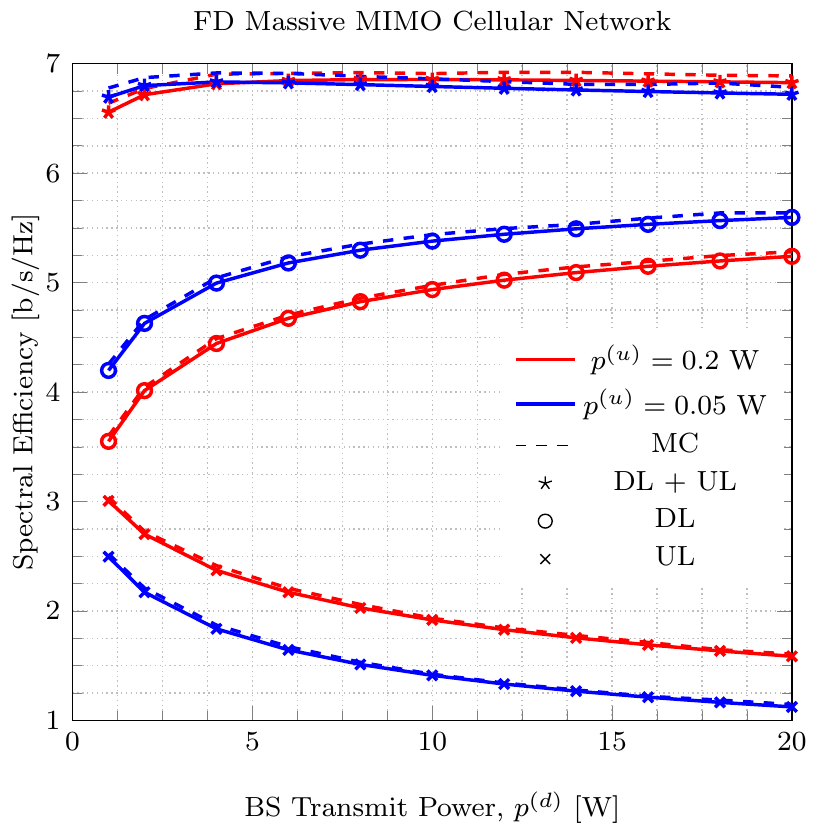}
\caption{System parameters are: $\lambda^{(d)} = \frac{4}{\pi}$ BSs/km$^{2}$, $N_{t} = 150$, $N_{r} = 50$, $\mathcal{U} = 4$, $p_{0} = -80$ dBm, $B = 20$ MHz, $N_{f} = 10$ dB, $\beta = 4$, $\psi = 1$, $\Omega = -100$ dB, $K = 0$.} 
\label{SEvsTxPowers}
\end{figure}

\begin{figure}[!t]
\centering
\includegraphics[scale=1]{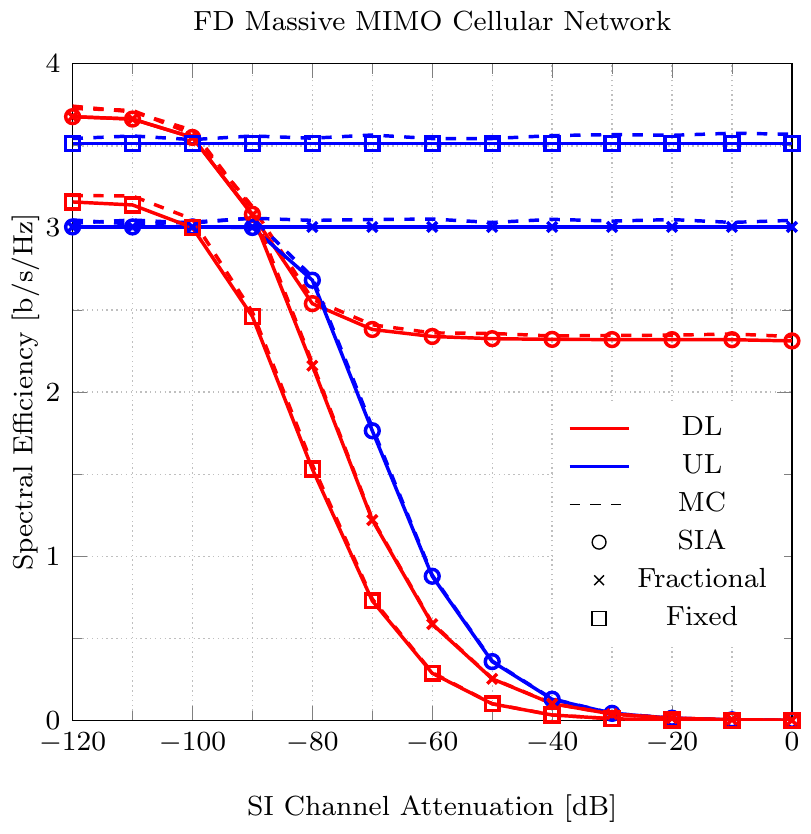}
\caption{System parameters are: $\lambda^{(d)} = \frac{4}{\pi}$ BSs/km$^{2}$, $N_{t} = 150$, $N_{r} = 50$, $\mathcal{U} = 4$, $p^{(d)} = 30$ dBm, $p^{(u)} = 23$ dBm, $p_{0} = -80$ dBm, $\frac{I_{\text{SI}}}{\sigma^2} = 25$ dB, $p^{(u)}_{f} = 23$ dBm, $B = 20$ MHz, $N_{f} = 10$ dB, $\beta = 4$, $\psi = 1$, $K = 0$.} 
\label{SEvsSIupndo}
\end{figure}

\subsubsection{Transmit Power Budget} It has been shown in \cite{6375940} that under perfect CSI, the transmit energy can be linearly conserved in the number of antennas. This is a contributing factor in tackling the UL rate bottleneck through massive MIMO, as was demonstrated in Fig. \ref{SEvsFDvsHD}. We depict the impact of different BS transmit power and MT maximum transmit power in Fig. \ref{SEvsTxPowers}. Intuitively, increasing $p^{(d)}$, or $p^{(u)}$, respectively improves the corresponding SE in the DL, or the UL; the relative gain however decreases for larger power budgets. Further, the DL and UL SEs are conflicting functions in the BS/MT transit powers. In general, we observe that a large difference in the DL/UL power levels is deteriorating to the overall performance in FD mode. Due to the large disparity in the DL/MT SEs, however, the direct optimization of the FD sum-rate places the focus on the DL, see Fig. \ref{SEvsTxPowers}. It can therefore be inferred that the optimization problem should be tackled by applying a weight to the UL SE. This is however beyond the scope of this work.   

\subsubsection{Uplink Power Control} The results presented so far were based on the conventional UL fractional power control mechanism defined in the existing LTE standards for HD cases. Next, we study the performance of the massive MIMO-enabled FD cellular network with different fixed (at maximum power), conventional, and proposed SIA fractional power control protocols under different SI channel attenuations in Fig. \ref{SEvsSIupndo}. It can be observed that the lack of self-interference-awareness in the case of fixed as well as fractional power allocation strategies means that the DL performance significantly suffers in the case of large SI channel attenuations. The proposed scheme can therefore serve as an important safe-guard mechanism for ensuring that a certain maximum SI level is not exceeded. 

\section{Summary}

We provided a theoretical framework using tools from stochastic geometry and point processes for the study of massive MIMO-enabled FD multi-cell multi-user cellular networks. The DL and UL SEs with linear ZF-SIN precoding, SIA fractional power control, and linear ZF receiver were characterized over Rician SI and Rayleigh intended and other-interference fading channels. The results highlighted the promising potential of massive MIMO towards unlocking the UL rate bottleneck in FD communication systems. On the other hand, the results demonstrated that the anticipated two-fold increase in SE of the massive MIMO-enabled FD cellular network overs its HD variant is only achievable in the asymptotic antenna region.

\bibliographystyle{IEEEtran}
\bibliography{IEEEabrv,myref}

\begin{thebibliography}{10}
\providecommand{\url}[1]{#1}
\csname url@samestyle\endcsname
\providecommand{\newblock}{\relax}
\providecommand{\bibinfo}[2]{#2}
\providecommand{\BIBentrySTDinterwordspacing}{\spaceskip=0pt\relax}
\providecommand{\BIBentryALTinterwordstretchfactor}{4}
\providecommand{\BIBentryALTinterwordspacing}{\spaceskip=\fontdimen2\font plus
\BIBentryALTinterwordstretchfactor\fontdimen3\font minus
  \fontdimen4\font\relax}
\providecommand{\BIBforeignlanguage}[2]{{%
\expandafter\ifx\csname l@#1\endcsname\relax
\typeout{** WARNING: IEEEtran.bst: No hyphenation pattern has been}%
\typeout{** loaded for the language `#1'. Using the pattern for}%
\typeout{** the default language instead.}%
\else
\language=\csname l@#1\endcsname
\fi
#2}}
\providecommand{\BIBdecl}{\relax}
\BIBdecl

\bibitem{6736746}
F.~Boccardi, R.~W. Heath, A.~Lozano, T.~L. Marzetta, and P.~Popovski, ``Five
  disruptive technology directions for {5G},'' \emph{IEEE Commun. Mag.},
  vol.~52, no.~2, pp. 74--80, Feb. 2014.

\bibitem{6375940}
H.~Q. Ngo, E.~G. Larsson, and T.~L. Marzetta, ``Energy and spectral efficiency
  of very large multiuser {MIMO} systems,'' \emph{IEEE Trans. Commun.},
  vol.~61, no.~4, pp. 1436--1449, Apr. 2013.

\bibitem{7401108}
P.~Patcharamaneepakorn, S.~Wu, C.~X. Wang, H.~Aggoune, M.~Alwakeel, X.~Ge, and
  M.~{Di Renzo}, ``Spectral, energy and economic efficiency of {5G} multi-cell
  massive {MIMO} systems with generalized spatial modulation,'' \emph{IEEE
  Trans. Veh. Technol.}, accepted 2016.

\bibitem{DBLP:journals/corr/GoyalGMP16}
S.~Goyal, C.~Galiotto, N.~Marchetti, and S.~Panwar, ``Throughput and coverage
  for a mixed full and half duplex small cell network,''
  \emph{arXiv:1602.09115}, 2016.

\bibitem{6353396}
M.~Duarte, C.~Dick, and A.~Sabharwal, ``Experiment-driven characterization of
  full-duplex wireless systems,'' \emph{IEEE Trans. Wireless Commun.}, vol.~11,
  no.~12, pp. 4296--4307, Dec. 2012.

\bibitem{6702851}
E.~Everett, A.~Sahai, and A.~Sabharwal, ``Passive self-interference suppression
  for full-duplex infrastructure nodes,'' \emph{IEEE Trans. Wireless Commun.},
  vol.~13, no.~2, pp. 680--694, Feb. 2014.

\bibitem{7263346}
M.~Chung, M.~S. Sim, J.~Kim, D.~K. Kim, and C.~B. Chae, ``Prototyping real-time
  full duplex radios,'' \emph{IEEE Commun. Mag.}, vol.~53, no.~9, pp. 56--63,
  Sept. 2015.

\bibitem{7041186}
D.~Kim, H.~Lee, and D.~Hong, ``A survey of in-band full-duplex transmission:
  From the perspective of {PHY} and {MAC} layers,'' \emph{IEEE Commun. Surveys
  Tuts.}, vol.~17, no.~4, pp. 2017--2046, Fourth Quart. 2015.

\bibitem{6847175}
D.~Nguyen, L.~N. Tran, P.~Pirinen, and M.~Latva-aho, ``On the spectral
  efficiency of full-duplex small cell wireless systems,'' \emph{IEEE Trans.
  Wireless Commun.}, vol.~13, no.~9, pp. 4896--4910, Sept. 2014.

\bibitem{DBLP:journals/corr/AtzeniK15}
I.~Atzeni and M.~Kountouris, ``Full-duplex {MIMO} small-cell networks:
  Performance analysis,'' \emph{arXiv:1504.04167}, 2015.

\bibitem{6963403}
G.~Zhang, K.~Yang, P.~Liu, and J.~Wei, ``Power allocation for full-duplex
  relaying-based {D2D} communication underlaying cellular networks,''
  \emph{IEEE Trans. Veh. Technol.}, vol.~64, no.~10, pp. 4911--4916, Oct. 2015.

\bibitem{7053957}
N.~Pappas, M.~Kountouris, A.~Ephremides, and A.~Traganitis, ``Relay-assisted
  multiple access with full-duplex multi-packet reception,'' \emph{IEEE Trans.
  Wireless Commun.}, vol.~14, no.~7, pp. 3544--3558, July 2015.

\bibitem{7403938}
M.~Mohammadi, B.~K. Chalise, H.~A. Suraweera, C.~Zhong, G.~Zheng, and
  I.~Krikidis, ``Throughput analysis and optimization of wireless-powered
  multiple antenna full-duplex relay systems,'' \emph{IEEE Trans. Commun.},
  vol.~64, no.~4, pp. 1769--1785, Apr. 2016.

\bibitem{DBLP:journals/corr/MohammadiST16}
M.~Mohammadi, H.~A. Suraweera, and C.~Tellambura, ``Full-duplex cloud-{RAN}
  with uplink/downlink remote radio head association,''
  \emph{arXiv:1602.08836}, 2016.

\bibitem{DBLP:journals/corr/AlAmmouriEAA15}
A.~AlAmmouri, H.~ElSawy, O.~Amin, and M.~Alouini, ``In-band full-duplex
  communications for cellular networks with partial uplink/downlink overlap,''
  \emph{arXiv:1508.02909}, 2015.

\bibitem{2016arXiv160400588S}
A.~Sadeghi, M.~Luvisotto, F.~Lahouti, S.~Vitturi, and M.~Zorzi, ``Statistical
  {QoS} analysis of full duplex and half duplex heterogeneous cellular
  networks,'' \emph{arXiv:1604.00588}, 2016.

\bibitem{7105653}
L.~Wang, F.~Tian, T.~Svensson, D.~Feng, M.~Song, and S.~Li, ``Exploiting full
  duplex for device-to-device communications in heterogeneous networks,''
  \emph{IEEE Commun. Mag.}, vol.~53, no.~5, pp. 146--152, May 2015.

\bibitem{2016arXiv160402602E}
A.~AlAmmouri, H.~ElSawy, and M.~S. Alouini, ``Flexible design for
  {$\alpha$}-duplex communications in multi-tier cellular networks,''
  \emph{IEEE Trans. Commun.}, vol.~64, no.~8, pp. 3548--3562, Aug. 2016.

\bibitem{6832464}
A.~Sabharwal, P.~Schniter, D.~Guo, D.~W. Bliss, S.~Rangarajan, and R.~Wichman,
  ``In-band full-duplex wireless: Challenges and opportunities,'' \emph{IEEE J.
  Sel. Areas Commun.}, vol.~32, no.~9, pp. 1637--1652, Sept. 2014.

\bibitem{7105646}
G.~Y. Li, M.~Bennis, and G.~Yu, ``Full duplex communications [guest
  editorial],'' \emph{IEEE Commun. Mag.}, vol.~53, no.~5, p.~90, May 2015.

\bibitem{6810573}
B.~Yin, M.~Wu, C.~Studer, J.~R. Cavallaro, and J.~Lilleberg, ``Full-duplex in
  large-scale wireless systems,'' in \emph{Proc. Asilomar Conf. Signals, Syst.
  and Comput. (ASILOMAR)}, Nov. 2013, pp. 1623--1627.

\bibitem{7428973}
K.~Min, S.~Park, Y.~Jang, T.~Kim, and S.~Choi, ``Antenna ratio for sum-rate
  maximization in full-duplex large-array base station with half-duplex
  multi-antenna users,'' \emph{IEEE Trans. Veh. Technol.}, accepted 2016.

\bibitem{DBLP:journals/corr/LimHC15}
Y.-G. Lim, D.~Hong, and C.-B. Chae, ``Performance analysis of self-interference
  cancellation methods in full-duplex large-scale {MIMO} systems,''
  \emph{arXiv:1508.02166}, 2015.

\bibitem{6983622}
S.~Huberman and T.~Le-Ngoc, ``Full-duplex {MIMO} precoding for sum-rate
  maximization with sequential convex programming,'' \emph{IEEE Trans. Veh.
  Technol.}, vol.~64, no.~11, pp. 5103--5112, Nov. 2015.

\bibitem{7492210}
J.~Kim, W.~Choi, and H.~Park, ``Beamforming for full-duplex multiuser {MIMO}
  systems,'' \emph{IEEE Trans. Veh. Technol.}, accepted 2016.

\bibitem{BaiS16}
J.~Bai and A.~Sabharwal, ``Large antenna analysis of multi-cell full-duplex
  networks,'' \emph{arXiv:1606.05025}, 2016.

\bibitem{7218556}
M.~A.~A. Khojastepour, K.~Sundaresan, S.~Rangarajan, and M.~Farajzadeh-Tehrani,
  ``Scaling wireless full-duplex in multi-cell networks,'' in \emph{Proc. IEEE
  Conf. Computer Commun. (INFOCOM)}, April 2015, pp. 1751--1759.

\bibitem{AtzeniK16a}
I.~Atzeni and M.~Kountouris, ``Full-duplex {MIMO} small-cell networks with
  interference cancellation,'' \emph{arXiv:1612.07289}, 2016.

\bibitem{psomas16}
C.~Psomas, M.~Mohammadi, I.~Krikidis, and H.~A. Suraweera, ``Directional
  antennas for interference management in full-duplex cellular networks,''
  \emph{arXiv:1602.02718}, 2016.

\bibitem{GeoffreyFD}
R.~Li, Y.~Chen, G.~Y. Li, and G.~Liu, ``Full-duplex cellular networks: It
  works!'' \emph{arXiv:1604.02852}, 2016.

\bibitem{7544500}
L.~Wang, K.~K. Wong, M.~Elkashlan, A.~Nallanathan, and S.~Lambotharan,
  ``Secrecy and energy efficiency in massive {MIMO} aided heterogeneous
  {C-RAN}: A new look at interference,'' \emph{IEEE J. Sel. Topics Signal
  Process.}, accepted, 2016.

\bibitem{6891254}
E.~Bj{\"o}rnson, J.~Hoydis, M.~Kountouris, and M.~Debbah, ``Massive {MIMO}
  systems with non-ideal hardware: Energy efficiency, estimation, and capacity
  limits,'' \emph{IEEE Trans. Inf. Theory}, vol.~60, no.~11, pp. 7112--7139,
  Nov. 2014.

\bibitem{CMTheorem}
S.~N. Chiu, D.~Stoyan, W.~S. Kendall, and J.~Mecke, \emph{Stochastic geometry
  and its applications}.\hskip 1em plus 0.5em minus 0.4em\relax John Wiley \&
  Sons, 2013.

\bibitem{6918448}
A.~Shojaeifard, K.~A. Hamdi, E.~Alsusa, D.~K.~C. So, and J.~Tang, ``A unified
  model for the design and analysis of spatially-correlated load-aware
  {HetNets},'' \emph{IEEE Trans. Commun.}, vol.~62, no.~11, pp. 1--16, Nov.
  2014.

\bibitem{7432156}
F.~Boccardi, J.~Andrews, H.~Elshaer, M.~Dohler, S.~Parkvall, P.~Popovski, and
  S.~Singh, ``Why to decouple the uplink and downlink in cellular networks and
  how to do it,'' \emph{IEEE Commun. Mag.}, vol.~54, no.~3, pp. 110--117, Mar.
  2016.

\bibitem{6516885}
T.~D. Novlan, H.~S. Dhillon, and J.~G. Andrews, ``Analytical modeling of uplink
  cellular networks,'' \emph{IEEE Trans. Wireless Commun.}, vol.~12, no.~6, pp.
  2669--2679, June 2013.

\bibitem{6919997}
H.~Y. Lee, Y.~J. Sang, and K.~S. Kim, ``On the uplink {SIR} distributions in
  heterogeneous cellular networks,'' \emph{IEEE Commun. Lett.}, vol.~18,
  no.~12, pp. 2145--2148, Dec. 2014.

\bibitem{7448861}
M.~D. Renzo and P.~Guan, ``Stochastic geometry modeling and system-level
  analysis of uplink heterogeneous cellular networks with multi-antenna base
  stations,'' \emph{IEEE Trans. Commun.}, vol.~64, no.~6, pp. 2453--2476, June
  2016.

\bibitem{7478073}
A.~Shojaeifard, K.~A. Hamdi, E.~Alsusa, D.~K.~C. So, J.~Tang, and K.~K. Wong,
  ``Design, modeling, and performance analysis of multi-antenna heterogeneous
  cellular networks,'' \emph{IEEE Trans. Commun.}, vol.~64, no.~7, pp.
  3104--3118, July 2016.

\bibitem{7805138}
A.~Shojaeifard, K.~K. Wong, M.~D. Renzo, G.~Zheng, K.~A. Hamdi, and J.~Tang,
  ``Self-interference in full-duplex multi-user {MIMO} channels,'' \emph{IEEE
  Commun. Lett.}, accepted 2017.

\bibitem{5706315}
S.~Parkvall, A.~Furuskar, and E.~Dahlman, ``Evolution of {LTE} toward
  {IMT}-advanced,'' \emph{IEEE Commun. Mag.}, vol.~49, no.~2, pp. 84--91, Feb.
  2011.

\bibitem{MDR-IntAware-arXiv}
F.~J. Martin-Vega, G.~Gomez, M.~C. Aguayo-Torres, and M.~Di~Renzo, ``Analytical
  modeling of interference aware power control for the uplink of heterogeneous
  cellular networks,'' \emph{arXiv:1601.03164}, 2016.

\bibitem{7277029}
A.~Shojaeifard, K.~A. Hamdi, E.~Alsusa, D.~K.~C. So, and J.~Tang, ``Exact
  {SINR} statistics in the presence of heterogeneous interferers,'' \emph{IEEE
  Trans. Inform. Theory}, vol.~61, no.~12, pp. 6759--6773, Dec. 2015.

\bibitem{adamchik}
V.~Adamchik and O.~Marichev, ``The algorithm for calculating integrals of
  hypergeometric type functions and its realization in {REDUCE} system,'' in
  \emph{Proc. Int. Symp. Symbolic and Algebraic Computation}.\hskip 1em plus
  0.5em minus 0.4em\relax ACM, 1990.

\bibitem{5407601}
K.~Hamdi, ``A useful lemma for capacity analysis of fading interference
  channels,'' \emph{IEEE Trans. Commun.}, vol.~58, no.~2, pp. 411--416, Feb.
  2010.

\bibitem{6516171}
M.~Di~Renzo, A.~Guidotti, and G.~Corazza, ``Average rate of downlink
  heterogeneous cellular networks over generalized fading channels: A
  stochastic geometry approach,'' \emph{IEEE Trans. Commun.}, vol.~61, no.~7,
  pp. 3050--3071, July 2013.

\bibitem{7511710}
T.~Bai and R.~W. Heath, ``Analyzing uplink {SINR} and rate in massive {MIMO}
  systems using stochastic geometry,'' \emph{IEEE Trans. Commun.}, accepted
  2016.

\bibitem{7031971}
E.~Bj{\"o}rnson, L.~Sanguinetti, J.~Hoydis, and M.~Debbah, ``Optimal design of
  energy-efficient multi-user {MIMO} systems: Is massive {MIMO} the answer?''
  \emph{IEEE Trans. Wireless Commun.}, vol.~14, no.~6, pp. 3059--3075, June
  2015.

\bibitem{7302534}
A.~He, L.~Wang, M.~Elkashlan, Y.~Chen, and K.~K. Wong, ``Spectrum and energy
  efficiency in massive {MIMO} enabled {HetNets}: A stochastic geometry
  approach,'' \emph{IEEE Commun. Lett.}, vol.~19, no.~12, pp. 2294--2297, Dec.
  2015.

\bibitem{1261332}
Q.~H. Spencer, A.~L. Swindlehurst, and M.~Haardt, ``Zero-forcing methods for
  downlink spatial multiplexing in multiuser {MIMO} channels,'' \emph{IEEE
  Trans. Signal Process.}, vol.~52, no.~2, pp. 461--471, Feb. 2004.

\end{thebibliography}

\appendices
\numberwithin{equation}{section}

\section{}
\label{app:ZF-SIN}

To jointly suppress the (residual) SI and multi-user interference in the DL, considering $\boldsymbol{G}_{l}$, $\boldsymbol{H}_{l}$, and $\boldsymbol{G}_{l,l}$ are full-rank and estimated without error, the conditions (i) $\boldsymbol{W}_{l} \boldsymbol{G}_{l,l} \boldsymbol{V}_{l} = \boldsymbol{0}_{\mathcal{U} \times \mathcal{U}}$, and (ii) $\boldsymbol{g}_{l,k} \boldsymbol{v}_{l,j} = 0, \; \forall j \neq k$ must be satisfied. The singular value decomposition (SVD) of $\boldsymbol{G}_{l,l}$ is given by $\pmb{\mathbbm{U}}_{l,l} \boldsymbol{\Sigma}_{l,l} \left[ \pmb{\mathbbm{V}}^{(1)}_{l,l} \pmb{\mathbbm{V}}^{(0)}_{l,l} \right]^{\dag}$ where $\pmb{\mathbbm{U}}_{l,l}$ holds the left singular vectors, $\boldsymbol{\Sigma}_{l,l}$ is the matrix of singular values, and $\pmb{\mathbbm{V}}^{(1)}_{l,l}$ and $\pmb{\mathbbm{V}}^{(0)}_{l,l}$ are  the right singular matrices of the non-zero and zero singular values, respectively \cite{1261332}. For SI nulling, the precoder is designed such that its column vectors are in the nullspace of $\boldsymbol{G}_{l,l}$ (i.e., $\boldsymbol{V}_{l} = \pmb{\mathbbm{V}}^{(0)}_{l,l}$) using ${\small \textbf{I}}_{N_{t}} - \boldsymbol{G}^{\dag}_{l,l} ( \boldsymbol{G}_{l,l} \boldsymbol{G}^{\dag}_{l,l} )^{-1} \boldsymbol{G}_{l,l}$. At the same time, to suppress multi-user interference in the DL, we design the proposed ZF-SIN precoder $\boldsymbol{V}_{l}$ using the pseudo-inverse of the projection channel matrix $\boldsymbol{\hat{G}}_{l} = \boldsymbol{G}_{l} ( {\small \textbf{I}}_{N_{t}} - \boldsymbol{G}^{\dag}_{l,l} ( \boldsymbol{G}_{l,l} \boldsymbol{G}^{\dag}_{l,l} )^{-1} \boldsymbol{G}_{l,l} )$.
\hfill $\blacksquare$

\section{}
\label{app:Power_Distribution_SIA_Rician}

The proposed SIA fractional power control mechanism c.d.f. can be expressed as
\begin{align}
\mathcal{F}_{p^{(u)}_{k,l}} \left( p \right) = \left( 1 - \mathcal{H} \left( p - p^{(u)} \right) \right) \mathscr{P} \left( \hat{p}^{(u)}_{k} \leq p \right) + \mathcal{H} \left( p - p^{(u)} \right) 
\end{align}
where $\hat{p}^{(u)}_{k} = \min \left(p_{0} d_{k,l}^{ \psi \beta},I_{\text{SI}} H^{-1}_{k,k} \right)$.
Hence, we can write
\begin{multline}
\mathcal{F}_{\hat{p}^{(u)}_{k}} \left( p \right) = \mathcal{F}_{p_{0} d_{k,l}^{ \psi \beta}} \left( p \right) + \mathcal{F}_{I_{\text{SI}} H^{-1}_{k,k}} \left( p \right) - \mathcal{F}_{p_{0} d_{k,l}^{ \psi \beta}} \left( p \right) \mathcal{F}_{I_{\text{SI}} H^{-1}_{k,k}} \left( p \right) \\ = \mathscr{P} \left( d_{k,l} \leq \left( \frac{p}{p_{0}} \right)^{\frac{1}{\psi \beta}} \right) + \mathscr{P} \left( H_{k,k} \geq \frac{I_{\text{SI}}}{p} \right)  - \mathscr{P} \left( d_{k,l} \leq \left( \frac{p}{p_{0}} \right)^{\frac{1}{\psi \beta}} \right) \mathscr{P} \left( H_{k,k} \geq \frac{I_{\text{SI}}}{p} \right).
\end{multline}
By considering the transmitter-receiver distance distribution 
\begin{align}
\mathscr{P} \left( d_{k,l} \leq \left( \frac{p}{p_{0}} \right)^{\frac{1}{\psi \beta}} \right) & = \int^{\left( \frac{p}{p_{0}} \right)^{\frac{1}{\psi \beta}}}_{0} 2 \pi \lambda^{(d)} r \exp \left( - \pi \lambda^{(d)} r^{2} \right) \diff r = 1 - \exp \left(- \pi \lambda^{(d)} \left( \frac{p}{p_{0}} \right)^{\frac{2}{\psi \beta}} \right), 
\end{align}
and the SI channel power gain non-central Chi-squared distribution
\begin{multline}
\mathscr{P} \left( H_{k,k} \geq \frac{I_{\text{SI}}}{p} \right) = \int^{+ \infty}_{\frac{I_{\text{SI}}}{p}} \frac{1 + K}{\Omega} \exp \left( - \left( K + \frac{(1 + K) h}{\Omega} \right) \right) I_0\left(2 \sqrt{ \frac{K (1 + K) h}{\Omega} }\right) \diff h \\ = Q_{1} \left(\sqrt{2 K}, \sqrt{\frac{2 (1 + K) I_{\text{SI}}}{p \Omega}}\right),
\end{multline}
we can obtain
\begin{align}
\mathcal{F}_{\hat{p}^{(u)}_{k}} \left( p \right) = 1 - \exp \left(- \pi \lambda^{(d)} \left( \frac{p}{p_{0}} \right)^{\frac{2}{\psi \beta}} \right) \left( 1 - Q_{1} \left(\sqrt{2 K}, \sqrt{\frac{2 (1 + K) I_{\text{SI}}}{p \Omega}}\right) \right).
\end{align}
We can write the following p.d.f. expression by differentiating the above
\begin{multline}
\mathcal{P}_{\hat{p}^{(u)}_{k}} \left( p \right) = \exp \left(-\pi \lambda^{(d)} \left(\frac{p}{p_{0}}\right)^{\frac{2}{\psi \beta}} \right) \mathlarger{\Biggr(} \frac{ 2 \pi \lambda^{(d)} }{\psi \beta p} \left(\frac{p}{p_{0}}\right)^{\frac{2}{\beta  \psi }} \left( 1-Q_1 \left(\sqrt{2 K},\sqrt{ \frac{2 (K+1) I_{\text{SI}} }{p \Omega}} \right) \right) \\ + \frac{(1+K) I_{\text{SI}} }{p^{2} \Omega} \exp \left( - \left( K + \frac{(1+K) I_{\text{SI}}}{p \Omega } \right)\right) \, _0\tilde{F}_1\left(;1;\frac{K (1+K) I_{\text{SI}}}{p \Omega }\right) \mathlarger{\Biggr)}.
\end{multline}
Hence, with some basic algebraic manipulations, we arrive at Lemma \ref{lem:Power_Distribution_SIA_Rician}.
\hfill $\blacksquare$

\section{}

\label{app:Power_Moments_SIA_SC}
Consider the case where the p.d.f. of the transmit power is
\begin{align}
\mathcal{P}_{p^{(u)}_{k,l}} \left( p \right) = \exp \left(- \pi \lambda^{(d)} \sqrt{\frac{p}{p_{0}}} \right) \Biggr[ \frac{ 2 \pi \lambda^{(d)} }{4 p} \sqrt{\frac{p}{p_{0}}} \left( 1 - \exp \left( - \frac{I_{\text{SI}}}{p \Omega} \right) \right) + \frac{I_{\text{SI}}}{p^{2} \Omega} \exp \left( - \frac{I_{\text{SI}}}{p \Omega} \right) \Biggr].
\label{eq:pdf_power_unbounded_Rayleigh}
\end{align} 
By utilizing the integral identities (where $n > 0$ and $\flat > n$)
\begin{align}
\int^{+ \infty}_{0} \exp \left( - x^{n} \right) x^{\flat - n} \diff x = \frac{\Gamma \left( \frac{1 - n - \flat}{n} \right)}{n}
\label{eq:InId_exp_pow1}
\end{align}
and
\begin{align}
\int^{+ \infty}_{0} \exp \left( - \sqrt{x} - \frac{1}{x} \right) x^{\flat - n} \diff x = \frac{1}{\sqrt{\pi}} \MeijerG[\Bigg]{3}{0}{0}{3}{n-\flat-1,0,\frac{1}{2}}{}{\frac{1}{4}},
\label{eq:InId_exp_pow2}
\end{align}
we can arrive at (\ref{eq:Power_Moments_SIA_Rayleigh_PL4}).

Next, consider the following transmit power p.d.f.  
\begin{align}
\mathcal{P}_{p^{(u)}_{k,l}} \left( p \right) = \frac{(1+K) I_{\text{SI}}}{p^{2} \Omega} \exp \left( - \left( K + \frac{(1+K) I_{\text{SI}}}{p \Omega} \right) \right) \, _0\tilde{F}_1\left(;1;\frac{K (1+K) I_{\text{SI}}}{p \Omega} \right).
\label{eq:pdf_power_unbounded_woCompensation}
\end{align}
We derive the following integral identity (where $\flat < 1$ and $n > \flat$)
\begin{align}
\int_0^{+ \infty } \exp \left( \frac{1}{p} \right) \, _0\tilde{F}_1\left(;1;\frac{1}{p}\right) p^{\flat -n} \, \diff p = L_{\flat - n + 1}(1) \Gamma (n-\flat -1).
\end{align}
Hence, by utilizing $\mathcal{L}_{m}(z) = \, _0\tilde{F}_{1}\left(-m;1;z\right)$, we can arrive at (\ref{eq:Power_Moments_SIA_Rician_woCompensation}). It should be noted that with no path-loss compensation, the first and beyond positive moments ($\flat \geq 1$) of $p^{(u)}_{k,l}$ are infinite. 

Finally, note the following unbounded transmit power p.d.f.  
\begin{align}
\mathcal{P}_{p^{(u)}_{k,l}} \left( p \right) = \frac{2 \pi \lambda^{(d)}}{\psi \beta p} \left(\frac{p}{p_{0}}\right)^{\frac{2}{\psi \beta}} \exp \left( - \pi \lambda^{(d)} \left(\frac{p}{p_{0}}\right)^{\frac{2}{\psi \beta}} \right).
\label{eq:pdf_power_unbounded_woSIA}
\end{align} 
By utilizing the integral identity (\ref{eq:InId_exp_pow1}), we can obtain (\ref{eq:Power_Moments_SIA_Rician_woSIA}). \hfill $\blacksquare$

\section{}
\label{app:Signals_MGFs}

Let $\Phi$, $\lambda$, and $\mathscr{E}$ denote the PPP, density, and exclusion region radius, respectively. Hence, $\mathcal{I} = \sum_{x \in \Phi} Q_{x} \| x \|^{- \beta}$ where $x$ and $Q_{x}$ are respectably the location and channel power gain of an arbitrary interferer with respect to a typical receiver at the origin. We proceed as follows
\begin{multline}
\mathcal{M}_{\mathcal{I}} (z) = \mathbb{E} \left\{ \exp \left( - z \sum_{x \in \Phi} Q_{x} \| x \|^{- \beta} \right) \right\}  
\overset{(i)}{=} \lim_{\mathscr{D} \rightarrow + \infty} \mathbb{E}_{\mathscr{N}} \left\{ \left( \mathbb{E}_{Q_{x},\| x \|} \left\{ \exp \left( - z Q_{x} \| x \|^{- \beta} \right) \right\} \right)^{\mathscr{N}} \right\} 
\\ \overset{(ii)}{=} \lim_{\mathscr{D} \rightarrow + \infty} \exp \left( \pi \lambda \left( \mathscr{D}^2 - \mathscr{E}^{2} \right) \left( \mathbb{E}_{Q_{x},\| x \|} \left\{ \exp \left( - z Q_{x} \| x \|^{- \beta} \right) \right\} - 1 \right) \right) \\ \overset{(iii)}{=} \lim_{\mathscr{D} \rightarrow + \infty} \exp \left( 2 \pi \lambda \mathbb{E}_{Q_{x}} \left( \int^{\mathscr{D}}_{\mathscr{E}} \mathcal{R} \left( \exp \left( - z Q_{x} \| \mathcal{R} \|^{- \beta} - 1 \right) \right) \diff \mathcal{R} \right) \right) \\ \overset{(iv)}{=} \exp \biggr( - \pi \lambda \mathbb{E}_{Q_{x}} \biggr\{ \biggr[ \Gamma \left( 1 - \frac{2}{\beta} \right) - \Gamma \biggr( 1 - \frac{2}{\beta} , z Q_{x} \mathscr{E}^{- \beta} \biggr) \biggr] \left( z Q_{x} \right)^{\frac{2}{\beta}} + \mathscr{E}^{2} \biggr( \exp \biggr( - z Q_{x}  \mathscr{E}^{- \beta} \biggr) - 1 \biggr) \biggr\} \biggr) 
\label{eq:generalInterference}
\end{multline}
where $(i)$ is written considering $\mathscr{N}$ conditional interferers are i.i.d. in a circular region of radius $\mathscr{D}$ ($> \mathscr{E}$) around the center with the limit as $\mathscr{D} \rightarrow + \infty$; $(ii)$ is from characterizing $\mathscr{N}$ as a Poisson random variable with expected value $\pi \lambda \left( \mathscr{D}^{2} - \mathscr{E}^{2} \right)$ and hence utilizing the Poisson identity $\mathbb{E} \left\{ \zeta^{\mathscr{N} } \right\} = \exp \left( \mathbb{E} \left\{ \mathscr{N} \right\} \left( \zeta - 1 \right) \right)$; $(iii)$ is written using the p.d.f. of the distance $\mathcal{R}$ (where $\mathscr{E} < \mathcal{R} < \mathscr{D}$) 
\begin{align}
\mathcal{P}_{\| x \|}(\mathcal{R}) = 
\frac{2 \mathcal{R}}{ \mathscr{D}^{2} - \mathscr{E}^{2}}\text{;}
\label{eq:Disss2}
\end{align}
$\left( iv \right)$ is obtained by invoking the integral identity (where $\alpha > 0$ and $\beta > 2$)
\begin{gather}
\mathbb{E}_{\| x \|} \left\{ \exp \left( - \alpha \| x \|^{- \beta} \right) \right\} = \frac{ 2 \alpha^{\frac{2}{\beta}}}{\beta \left(\mathscr{D}^2-\mathscr{E}^{2} \right)} \left[ \Gamma \left(- \frac{2}{\beta} , \frac{\alpha}{\mathscr{D}^{\beta}} \right)-\Gamma \left(- \frac{2}{\beta},\frac{\alpha}{\mathscr{E}^{\beta}} \right) \right],
\label{eq:InteIdent}
\end{gather}
and taking the limit as $\mathscr{D} \rightarrow + \infty$.

To proceed, the distribution of the channel power gain should be specified. Here, we consider the general case where $Q_{x} \sim \Gamma(U,V)$, which can be used to capture a wide range of MIMO setups \cite{6516171}. Hence, through utilizing the following integral identities (where $\alpha > 0$ and $\beta > 2$)
\begin{align}
\mathbb{E}_{Q_{x}} \left\{ Q_{x}^{\frac{2}{\beta}} \right\} & = \frac{\Gamma \left(U+\frac{2}{\beta }\right)}{\Gamma (U)} V^{\frac{2}{\beta}},
\label{eq:channelIden1}
\end{align}
\begin{align}
\mathbb{E}_{Q_{x}} \left\{ \Gamma \left(1-\frac{2}{\beta},\alpha Q_{x} \right) \right\} = \frac{\Gamma \left(U-\frac{2}{\beta }+1\right)}{\left( \alpha V \right)^{U} \Gamma (U+1)} \, _2F_1\left(U,U-\frac{2}{\beta }+1;U+1;-\frac{1}{\alpha V}\right),
\label{eq:channelIden2}
\end{align}
and
\begin{align}
\mathbb{E}_{Q_{x}} \left\{ \exp \left( - \alpha Q_{x} \right) \right\} = \frac{1}{(1 + \alpha V )^{U}},
\label{eq:channelIden3}
\end{align}
we can arrive at
\begin{multline}
\mathcal{M}_{\mathcal{I}} (z) = \exp \mathlarger{\Biggr(} - \pi  \lambda \mathlarger{\Biggr[} -\mathscr{E}^{2} + \frac{\mathscr{E}^{2}}{\left(z V \mathscr{E}^{-\beta} +1\right)^{U}} + \frac{ \Gamma \left(1-\frac{2}{\beta} \right) \Gamma \left(U+\frac{2}{\beta} \right)}{\Gamma \left( U \right)} \left( z V \right)^{\frac{2}{\beta}} \\ - \frac{U \beta}{U \beta + 2} \frac{\mathscr{E}^{ U \beta +2}}{\left( z V \right)^{U}} \, _2F_1\left(U+1,U+\frac{2}{\beta};U+\frac{2}{\beta }+1;- \frac{\mathscr{E}^{\beta}}{z V} \right) \mathlarger{\Biggr]} \mathlarger{\Biggr)}.
\label{eq:generalInterferenceGamma}
\end{multline}
\hfill $\blacksquare$

\section{}
\label{app:SE}

The DL and UL SE expressions respectively provided in (\ref{eq:SE_DL}) and (\ref{eq:SE_UL}) can be obtained directly through applying the m.g.f.-based approach in \cite{5407601} to a stochastic geometry-based setting.

Consider the cases where the DL and UL SEs are respectively simplified to
\begin{multline}
\mathcal{S}_{d,f} = \log_{2} (e) \int^{+ \infty}_{0} \int^{+ \infty}_{0} \int^{+ \infty}_{0} \frac{2 \pi \lambda^{(d)} r p^{(d)}}{(1 + z p \Omega) \left( r^4 + z p^{(d)} \right)} \\ \exp \left(-\pi \lambda^{(d)} \left( \frac{\pi}{2} \sqrt{z p} + \sqrt{z p^{(d)}} \arctan \left( \frac{\sqrt{z p^{(d)}}}{r^{2}} \right) + r^{2} \right) \right) \mathcal{P}_{p^{(u)}_{k,l}}  \diff z \diff r \diff p
\end{multline}
and
\begin{multline}
\mathcal{S}_{u,f} = \log_{2} (e) \int^{+ \infty}_{0} \int^{+ \infty}_{0} \int^{+ \infty}_{0} \frac{2 \pi \lambda^{(d)} r p}{1 + z p^{(d)} \Omega) \left( r^4 + z p \right)} \\ \exp \left(-\pi \lambda^{(d)} \left( \frac{\pi}{2} \sqrt{z p^{(d)}} + \sqrt{z p} \arctan \left( \frac{\sqrt{z p}}{r^{2}} \right) + r^{2} \right) \right) \mathcal{P}_{p^{(u)}_{k,l}}  \diff z \diff r \diff p. 
\end{multline}
Utilizing, in the order given, substitution with $\frac{z}{p^{(d)}} \rightarrow u^4$, conversion from Cartesian to polar coordinates with $u \rightarrow \mathscr{R} \sin (\mathscr{T})$ and $r \rightarrow \mathscr{R} \cos (\mathscr{T})$ (with Jacobian $\mathscr{R}$), the integral identity
\begin{align}
\int_0^{+ \infty} \frac{\mathscr{R}}{ 1 + \mathscr{R}^{4}} \exp \left( - \mathscr{R}^2 \right) \diff \mathscr{R} = \frac{1}{2} \left( \sin \left( 1 \right) \mathscr{C} \left( 1 \right) + \left(\frac{\pi}{2} -  \cos \left( 1 \right) \mathscr{S} \left( 1 \right) \right) \right),
\end{align} 
and substitution with $\mathscr{T} \rightarrow \arctan \sqrt{s}$, we can arrive at (\ref{eq:SE_SC}). The same process can be applied in the case of complete SI removal but with the reduced integral identity (\ref{eq:pdf_power_unbounded_Rayleigh}) to obtain (\ref{eq:SE_SC_noSI}). Finally, we can respectively arrive at (\ref{eq:SE_DL_SC_woComp})-(\ref{eq:SE_UL_SC_woComp}) using
\begin{align}
\int_0^{+ \infty} \frac{1}{p \left(p+\sqrt{p}\right)} \exp \left( -\frac{1}{p} \right)\diff p = \sqrt{\pi} + \frac{1}{e} \text{Ei}(1) - \frac{\pi}{e} \text{erfi}(1)  
\end{align}
and at (\ref{eq:SE_DL_SC_woSI})-(\ref{eq:SE_UL_SC_woSI}) using
\begin{align}
\int_0^{+ \infty} \frac{1}{p + \sqrt{p}} \exp \left( -\sqrt{p} \right) \, dp = - 2 \exp (1) \text{Ei}(-1).
\end{align}
\hfill $\blacksquare$

\section{}
\label{app:FDvsHD}

By computing the special case of $2 \left( \mathcal{S}_{d,f} + \mathcal{S}_{u,f} \right)$ and $\left( \mathcal{S}_{d,h} + \mathcal{S}_{u,h} \right)$ using the approach in Appendix \ref{app:SE}, we can readily obtain (\ref{eq:FD_SE_SC_EXACT}) and (\ref{eq:HD_SE_SC_EXACT}), respectively. 

Utilizing the approximation (where $s \geq 0$)
\begin{align}
\arctan(s)  =  \arcsin \left( \sqrt{\frac{s^2}{1 + s^2}} \right)  \geq \frac{s}{1 + s},
\end{align}
the integral identity (where $\alpha > \frac{1}{4}$)
\begin{multline}
\int_0^{\infty } \frac{1 + s}{\left( 1 + s^2 \right) \left( \alpha s^2 + s + 1 \right)} \, \diff s = \frac{1}{\sqrt{4 \alpha-1} \left(\alpha^2-2 \alpha + 2 \right)} \biggr( \pi  \left(\alpha^2-\frac{\alpha}{2} \left(\sqrt{4 \alpha-1}+3\right) +\sqrt{4 \alpha-1}\right) \\ +\frac{\alpha}{2} \sqrt{4 \alpha-1} \log (\alpha)-\alpha (2 \alpha-3) \arccot \left(\sqrt{4 \alpha-1}\right) \biggr),
\end{multline}
and basic algebraic manipulation, we derive (\ref{eq:FD_SE_APPROX}) and (\ref{eq:HD_SE_APPROX}).
\hfill $\blacksquare$

\section{}
\label{app:optX_FDvsHD}
We proceed by defining the super-level set of $\frac{\mathbbm{S}_{f}}{\mathbbm{S}_{h}}$ in $x$ as
\begin{align}
\mathcal{L} = \left\{ x \in \text{{\small \textbf{dom}} } \frac{\mathbbm{S}_{f}}{\mathbbm{S}_{h}} \Big| \frac{\mathbbm{S}_{f}}{\mathbbm{S}_{h}} \geq \mathscr{K} \right\}.
\end{align}
For the case in which $\mathscr{K} < 0$, the total SE gain of FD over HD is always positive, hence, there are no points on the counter, $\left( \frac{\mathbbm{S}_{f}}{\mathbbm{S}_{h}} \right)^{*} = \mathscr{K}$. 
On the other hand, for $\mathscr{K} > 0$, we can write 
\begin{align}
\frac{\mathbbm{S}_{f}}{\mathbbm{S}_{h}} = \frac{2 \left( \mathcal{S}_{d,f} + \mathcal{S}_{u,f} \right)}{\mathcal{S}_{d,h} + \mathcal{S}_{u,h}} = \frac{\Psi \left( \frac{8}{\pi \left( 1 + \sqrt{x} \right)}-1 \right) + \Psi \left( \frac{8}{\pi \left( 1 + \frac{1}{\sqrt{x}} \right)}-1 \right)}{\Psi \left( \frac{8}{\pi}-1 \right)}
\end{align} 
with
\begin{align}
\mathcal{L} \equiv \Psi \left( \frac{8}{\pi}  - 1 \right) \mathscr{K} - \left( \Psi \left( \frac{8}{\pi \left( 1 + \sqrt{x} \right)}-1 \right) + \Psi \left( \frac{8}{\pi \left( 1 + \frac{1}{\sqrt{x}} \right)}-1 \right) \right) \leq 0.
\end{align} 
The joint HD DL/UL SE in the above is affine in $x$. It can be shown that the second derivative of the FD mode SE function, $\frac{\diff^{2}}{\diff x^{2}} \left( 2 \left( \mathbbm{S}_{d,f} + \mathbbm{S}_{d,f} \right) \right)$, is always positive for $\left( \frac{\pi}{\pi - 8} \right)^{2} < x < 1$, always negative for $1 < x < \left( \frac{\pi - 8}{\pi} \right)^{2}$, and zero for $x = 1$. Consequently, $\mathcal{L}$ is convex in $x$ and the objective function in (\ref{eq:opt_prob_power_simplified}) is strictly quasi-concave in $x$, with a maximum point at $x^{*} = 1$. 
\hfill $\blacksquare$

\end{document}